\documentclass[%
reprint,
superscriptaddress,
amsmath,amssymb,
aps,
longbibliography,
prb,
]{revtex4-2}

\usepackage{graphicx}
\graphicspath{{../figs/}}
\usepackage[rgb]{xcolor}
\definecolor{dgreen}{rgb}{0.0,0.7,0.0}

\usepackage{braket}
\usepackage{lipsum}
\usepackage{siunitx}

\DeclareUnicodeCharacter{2212}{-}

\usepackage{dcolumn}
\newcolumntype{d}[1]{D{.}{.}{#1}}

\usepackage{bm}

\usepackage{layouts} 

\usepackage{subcaption}
\usepackage[justification=Justified]{caption}
\usepackage{pifont}

\usepackage[utf8]{inputenc}

\newcommand{\MAPI}{MAPbI$_3$}
\newcommand{\FAPI}{FAPbI$_3$}
\newcommand{\MAFAPI}{MA$_{x}$FA$_{1-x}$PbI$_{3}$}
\newcommand{\MAFAPIxy}[2]{MA$_{#1}$FA$_{#2}$PbI$_3$}
\newcommand{\eh}{$\rm{e}$--$\rm{h}$}

\newcommand{\CsTenFAPIBr}{FA$_{0.9}$C\lowercase{s}$_{0.1}$P\lowercase{b}I$_{2.8}$B\lowercase{r}$_{0.2}$}

\newcommand{\tauRX}{{\tau_{\rm R}^{\rm X}}}
\newcommand{\tauReh}{{\tau_{\rm R}^{\rm eh}}}
\newcommand{\tauSX}{{\tau_{\rm s}^{\rm X}}}
\newcommand{\tauSeh}{{\tau_{\rm s}^{\rm eh}}}
\newcommand{\initPLratio}{{I_{\rm X}(0)/I_{\rm eh}(0)}}

\newcommand{\Poo}{P_{\rm oo}}
\newcommand{\PooX}{\Poo^{\rm X}}
\newcommand{\Pooeh}{\Poo^{\rm eh}}

\newcommand{\WcmSq}{W/cm$^{2}$}

\usepackage{float}
\usepackage{blindtext}
\usepackage{soul}
\usepackage[unicode, colorlinks=blue, urlcolor=blue, filecolor=blue]{hyperref}

\begin{document}

\preprint{APS/123-QED}

\title{Spin dynamics of excitons and carriers in mixed-cation \MAFAPI{} perovskite crystals: alloy fluctuations probed by optical orientation}

\author{B. F. Gribakin}
\email{bgribakin@gmail.com}

\author{N. E. Kopteva}%
\email{natalia.kopteva@tu-dortmund.de}

\author{D. R. Yakovlev}
\affiliation{%
Experimentelle Physik 2, Technische Universit{\"a}t Dortmund, 44227 Dortmund, Germany}

\author{I. A. Akimov}
\affiliation{%
Experimentelle Physik 2, Technische Universit{\"a}t Dortmund, 44227 Dortmund, Germany}

\author{I. V. Kalitukha}
\affiliation{%
Experimentelle Physik 2, Technische Universit{\"a}t Dortmund, 44227 Dortmund, Germany}

\author{B.~Turedi}
\affiliation{Laboratory of Inorganic Chemistry, Department of Chemistry and Applied Biosciences,  ETH Z\"{u}rich, CH-8093 Z\"{u}rich, Switzerland}
\affiliation{Laboratory for Thin Films and Photovoltaics, Empa-Swiss Federal Laboratories for Materials Science and Technology, CH-8600 D\"{u}bendorf, Switzerland}

\author{M.~V.~Kovalenko}
\affiliation{Laboratory of Inorganic Chemistry, Department of Chemistry and Applied Biosciences,  ETH Z\"{u}rich, CH-8093 Z\"{u}rich, Switzerland}
\affiliation{Laboratory for Thin Films and Photovoltaics, Empa-Swiss Federal Laboratories for Materials Science and Technology, CH-8600 D\"{u}bendorf, Switzerland}

\author{M. Bayer}
\affiliation{%
Experimentelle Physik 2, Technische Universit{\"a}t Dortmund, 44227 Dortmund, Germany}
\affiliation{%
Research Center FEMS, Technische Universit\"at Dortmund, 44227 Dortmund, Germany}

\date{\today}

\begin{abstract}
Optical spin orientation measured by time-resolved photoluminescence provides a powerful tool to probe the spin dynamics of excitons and charge carriers in perovskite semiconductors. The impact of alloy fluctuations on the spin dynamics of mixed-cation \MAFAPI{} perovskite single crystals is studied here experimentally. The optical orientation is measured under nonresonant excitation for crystals with $x = 0.1$, $0.4$, and $0.8$ at cryogenic temperatures and compared with data on \MAPI{} crystals. The high degree of exciton optical orientation  of  $75-80$\% for $x = 0.1$ and $0.8$  reduces to about 60\% for $x = 0.4$. A similar trend is observed for the carrier spin optical orientation.  This behavior is attributed to enhanced  scattering of free excitons and carriers in the alloys with increased compositional and structural disorder.  From the Larmor spin precession measured from spin dynamics in an external  magnetic field applied in the Voigt geometry, the electron and hole $g$-factors are evaluated. Their dependence on the band gap energy in \MAFAPI{} crystals follows the universal trend previously established for lead halide perovskites.

\end{abstract}

\maketitle

\section{Introduction} 
\label{sec:introduction}

The research interest in the optical properties of lead-halide perovskite semiconductors continues to grow because of their outstanding photovoltaic~\cite{jena2019} and optoelectronic properties~\cite{Vinattieri2021_book,Vardeny2022_book}. Lead-halide perovskite semiconductors have the chemical formula APbX$_3$. Substitution of the type of anion (X$^{-}$ = I$^-$, Br$^-$ or Cl$^-$) strongly affects the band gap energy~\cite{protesescu_kovalenko2015nanolett_CsPbX3_bandgap_tuning}, whereas cation (A$^{+}$ = MA$^{+}$, FA$^{+}$ or Cs$^{+}$) substitution has a large effect on the structural properties~\cite{wright_herz2016ncomms_FAPbI3Br3_MAPbI3Br3_phonos, mohanty_sarma2019acsenlett_FAMAPbI3_phase_diagram}, governed by the size and dipole moment of the cation. Partial substitution of FA$^{+}$ by Cs$^{+}$ and/or MA$^{+}$ is commonly used to suppress the degradation of \FAPI{} to the yellow $\delta$-phase, which occurs in pure \FAPI{} due to the metastability of the cubic $\alpha$-phase~\cite{lee_park2015advenmat_CsFAPbI3_stable, li_zhu2016chemmat_CsFAPbI3_stable}.
Indeed, photovoltaic materials based on \MAFAPI{} films appear to be promising for combining the stability of \MAPI{} with the more application-suitable band gap of \FAPI{}. Multiple reports indicate that films with $x \sim 0.5$ deliver optimal photovoltaic performance across the gamut of materials with $0 < x < 1$~\cite{duan_zhou2018optmat_FAMAPI_PV, yang_xiao2019pccp_FAMAPI_PV, luo_xiao2020jelmat_FAMAPI_PV}.

The structure of mixed-cation \MAFAPI{} perovskites has been studied by various techniques, including single-crystal X-ray diffraction and Raman spectroscopy~\cite{weber_weller2016jmatchemA_FAMAPbI3_phase_transitions, mohanty_sarma2019acsenlett_FAMAPbI3_phase_diagram, francisco-lopez_goni2020jphyschemC_FAMAPbI3_phase_diagram}. These studies show that solution-grown mixed-cation \MAFAPI{} forms well-structured crystals with good optical properties. However, there is some disagreement in how the observed crystal phases are identified, which suggests that the temperature vs. MA content phase diagram depends either on the experimental conditions, or on the crystals themselves, i.\,e. on the growth parameters.

Optical spectroscopy using polarized light in magnetic field probes the spin-dependent properties of semiconductors, providing rich information about crystal structure and properties through the exciton and charge carrier recombination and spin dynamics. Optical spin orientation is an established technique for obtaining detailed information on the spin dynamics in semiconductors~\cite{opticalOrientation1984,dyakonov2017spin_ch1}.

\begin{figure}[t]
    \centering
    \includegraphics[width=0.45\textwidth]{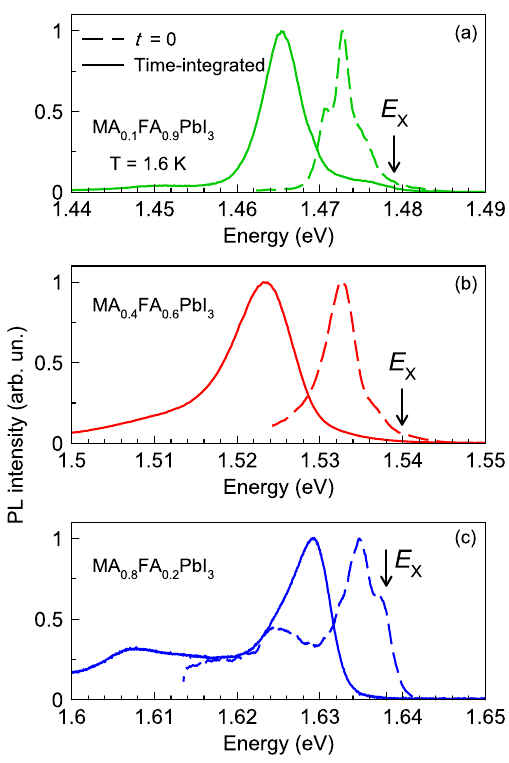}
    \caption{    
     Time-integrated photoluminescence spectra of the studied \MAFAPI{} crystals measured at $T = 1.6$~K using the laser photon energy $E_{\rm exc} = 1.77$~eV for excitation with power density $P = 0.5$~\WcmSq{} for $x= 0.1$ (a), $x= 0.4$ (b), and $x= 0.8$ (c). Photoluminescence spectra measured at the moment of pulse arrival are shown by the dashed lines. $E_\text{X}$ denotes the exciton resonance.
    }
    \label{fig:TIPL}
\end{figure}

\begin{figure}[t]
    \centering
    \includegraphics[]{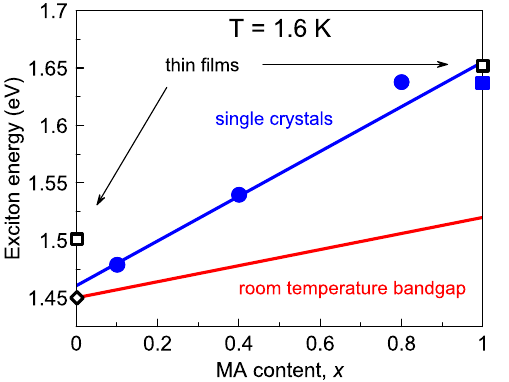}
    \caption{
    Exciton energies $E_{\rm X}$ in \MAFAPI{} crystals at $T = 1.6$~K, measured in this work (blue dots) and in Ref.~\cite{kopteva_bayer2025PRB_MAPbI3_OO} (blue square), plotted against MA content. The blue line is a linear fit of these data. Data for polycrystalline films at $T = 2$~K are shown by black empty squares (Ref.~\cite{galkowski_nicholas2016enenvsci_FAPI_MAPI_Eg_Ex}, magneto-optical band gap measurement) and empty diamond (Ref.~\cite{fang_anoniettaLoi2016lightSciAppl_FAPI_film_spectra}, estimate of band gap from optical spectra).
    The solid red line is a linear fit of room-temperature band gap measurements as reported in Ref.~\cite{weber_weller2016jmatchemA_FAMAPbI3_phase_transitions}.
    The exciton binding energy is about $15$~meV for both \FAPI{} and \MAPI{} at $T=2$~K~\cite{galkowski_nicholas2016enenvsci_FAPI_MAPI_Eg_Ex}. 
   }
    \label{fig:Eg_vs_x}
\end{figure}

\begin{figure*}[t]
    \centering
    \includegraphics[]{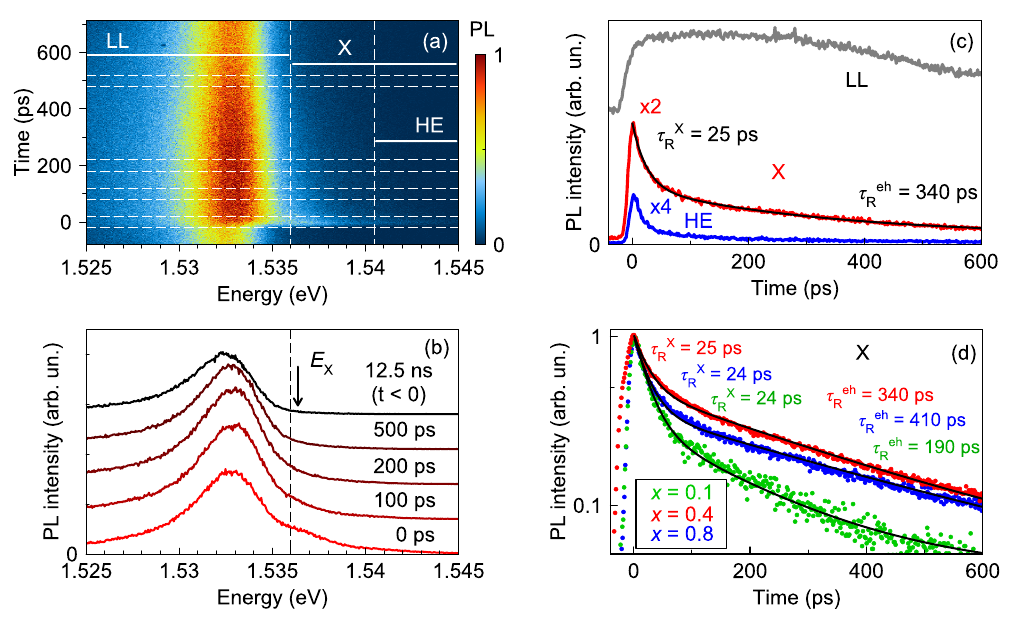}
    \caption{     
    Photoluminescence dynamics measured at $T = 1.6$~K in a \MAFAPIxy{0.4}{0.6} crystal. (a) Spectrally-resolved photoluminescence dynamics excited by $100$~fs pulses with excitation at $E_{\rm exc} = 1.70$~eV using $P = 0.5$~\WcmSq{} laser fluence. The horizontal dashed lines mark the temporal ranges used for the time-integrated data presented in panel (b).
    The vertical dashed lines mark the the high- and low-energy boundaries of the spectral ranges for the long-lived dynamics (LL, $1.525-1.536$~eV), exciton (X, $1.536-1.545$~eV), and the high energy range (HE, $1.5405-1.545$~eV) (shown by solid lines) used for the time-resolved data in panels (c, d).  
    (b) PL spectra measured in a 40~ps time window at different delays as marked by the horizontal dashed lines in panel (a). The signal at $t < 0$ is measured at $t = -50$~ps and is equivalent to $t \approx 12.5$~ns. The spectra are shifted vertically for clarity.
    (c) Photoluminescence dynamics, spectrally integrated across the LL, X, and HE ranges.
    (d) Recombination dynamics in \MAFAPI{} crystals measured with $E_{\rm exc} = 1.77$~eV and $P = 0.5$~\WcmSq{}, averaged over the X spectral range (see Section~\ref{SI:trpl} in the Supplementary Information for the exact ranges in the other samples). The black lines are biexponential fits.}
    \label{fig:TRPL_x04}
\end{figure*}

Recently, a giant degree of optical orientation was demonstrated for bulk \MAPI{} and \CsTenFAPIBr{} crystals, reaching up to 85\% for excitons and 50–60\% for charge carriers~\cite{kopteva_bayer2024advSci_FAPbI3_OO, kopteva_bayer2025PRB_MAPbI3_OO}. Remarkably, such high polarization persists even for optical detunings of several hundred meV from the exciton resonance to higher energies. This indicates that at cryogenic temperatures, the spin relaxation is inefficient not only for thermalized excitons and carriers, but also for those with excess kinetic energy.

It is, therefore, of interest to examine whether these features are preserved in mixed-cation \MAFAPI{} crystals. In these materials, the inherent disorder arising from alloy fluctuations may strongly enhance exciton–phonon scattering, particularly during the final stage of energy relaxation involving acoustic phonons. At the same time, the alloy fluctuation potential can localize carriers and excitons, at least at cryogenic temperatures. Localization suppresses spin relaxation mechanisms typical for mobile particles, while simultaneously it enhances the importance of the hyperfine interaction between carrier spins and nuclear spins~\cite{kudlacik2024optical,kotur2025nucleiFAPI}.

The impact of alloy fluctuations on the exciton and carrier spin dynamics remains relatively unexplored in perovskite semiconductors, although it is well known to be significant in conventional semiconductor alloys such as (Al,Ga)As or Cd(S,Se)~\cite{opticalOrientation1984,dyakonov2017spin}. 
Recent studies of mixed lead halide perovskites demonstrated important role of band gap fluctuations, which leads to pronounced optical response from localized excitons with extended coherence times up to 60~ps~\cite{Grisard2023}. 
In addition, alloy fluctuations may locally reduce the crystal symmetry, for example by breaking the spatial inversion symmetry. Such symmetry reduction could activate the Dyakonov–Perel spin relaxation mechanism and thus manifest itself as an acceleration of the spin dynamics. Notably, the ideal perovskite lattice possesses inversion symmetry, which suppresses the Dyakonov–Perel mechanism, although various routes to inversion symmetry breaking have been widely discussed in literature.
In this context, time-resolved photoluminescence (TRPL) measurements offer direct access to the spin dynamics on the recombination timescale, enabling simultaneous probing of excitons and spatially separated electron-hole (\eh{}) pairs.

In this paper, we report time-resolved optical orientation experiments on a set of \MAFAPI{}  crystals with $x = 0.1$, $0.4$, and $0.8$ in order to examine the impact of alloy fluctuations on the spin dynamics. The optical orientation degree is measured for various detunings of the laser excitation energy from the exciton resonance, and as a function of the excitation power density and crystal temperature. Experiments in transverse magnetic field reveal spin beats of electrons and holes, which allows us to evaluate their Land\'e $g$-factors.

\section{Results and Discussion}

\subsection{Optical properties}
\label{sec:opt_prop}

\begin{table*}[htbp]
\renewcommand{\arraystretch}{1.5}
\caption{
Recombination dynamics and spin parameters of excitons and carriers in \MAFAPI{}  crystals at $T=1.6$~K. The data are given for $E_{\rm exc}=1.77$~eV and  $P = 0.5$~W/cm$^2$. Data on \MAPI{}~\cite{kopteva_bayer2025PRB_MAPbI3_OO} ($E_{\rm exc}=1.771$~eV) and \CsTenFAPIBr{} crystals~\cite{kopteva_bayer2024advSci_FAPbI3_OO} ($E_{\rm exc}=1.669$~eV) are included for comparison.
}
\begin{ruledtabular}
\begin{tabular}{ccccccccc} 
  &  $E_{\rm X}$~(eV) & $\tauRX$~(ps) & $\tauReh$~(ps) & $\PooX(0)$ & $\Pooeh(0)$ & $g_{e}$ & $g_{h}$ \\ \hline
 \MAFAPIxy{0.1}{0.9} &  $1.479$ & 24 & 190 & $0.75$ & 0.55 & $+3.71$ & $-1.34$ \\
 \MAFAPIxy{0.4}{0.6} &  $1.540$ & 25 & 340 & $0.60$ & 0.35 & $+3.27$ & $-1.02$ \\
 \MAFAPIxy{0.8}{0.2} &  $1.638$ & 24 & 410 & $0.80$ & 0.70 & $+2.86$ & $-0.52$ \\
 \MAPI{}~\cite{kopteva_bayer2025PRB_MAPbI3_OO} &  $1.636$ & 15 & 520 & $0.85$ & 0.50 & $+2.83$ & $-0.54$ \\
 \CsTenFAPIBr{}~\cite{kopteva_bayer2024advSci_FAPbI3_OO} &  $1.506$ & 55 & 840 & $0.85$ & $0.60$ & $+3.48$ & $-1.15$ \\
\end{tabular}
\end{ruledtabular}
\label{table:parameters_summary}
\end{table*}

The samples under study are single \MAFAPI{} crystals with $x = 0.1$, $0.4$, and $0.8$  grown by the inverse temperature crystallization method. Their thickness is about 30~$\mu$m. More details on the samples and experimental conditions are given in Section~\ref{sec:experimental}.

\begin{figure*}[t]
    \centering
    \includegraphics[]{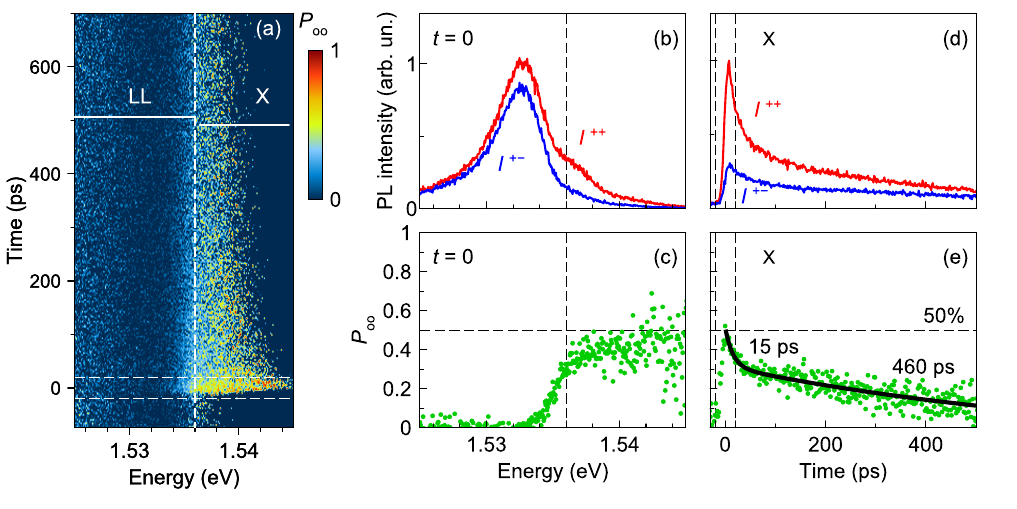}
    \caption{Optical orientation of excitons and carriers in a \MAFAPIxy{0.4}{0.6} crystal at $T = 1.6$~K, measured using $\sigma^+$ polarized excitation. $E_{\rm exc} = 1.70$~eV and $P= 0.5$~W/cm$^2$ at $T = 1.6$~K.
    (a) Spectrally-resolved dynamics of the optical orientation degree. The dashed lines show the temporal and spectral ranges used to average the data presented in panels (b,c) and (d,e), respectively.
    (b) PL spectra detected in $\sigma^+$ and $\sigma^-$ circular polarization at the moment of pulse arrival.
    (c) Spectral dependence of the optical orientation degree at $t=0$ calculated from the data in panel (b). The horizontal dashed line gives the maximum optical orientation degree.
    (d) Spectrally-integrated ($1.536-1.545$~eV) PL dynamics detected in $\sigma^+$ and $\sigma^-$ circular polarization.
    (e) Dynamics of the optical orientation degree calculated from the data in panel (d). The black line is a biexponential fit giving estimates for the exciton and \eh{}-pair initial optical orientation degrees of $\PooX(0) = 0.50$ and $\Pooeh(0) = 0.35$ and their decay times of 15~ps and 460~ps, respectively.
     }
    \label{fig:TRPL_P_OO}
\end{figure*}

\begin{figure*}[t]
    \centering
    \includegraphics[]{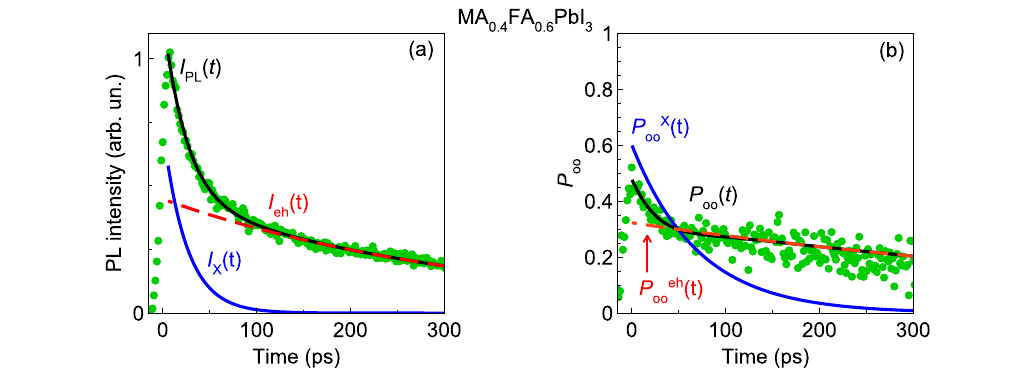}
    \caption{
    (a)  Experimental PL dynamics (dots) in a \MAFAPIxy{0.4}{0.6} crystal measured with $E_{\rm exc} = 1.70$~eV and $P= 0.5$~W/cm$^2$ at $T = 1.6$~K.
    The black line shows a biexponential fit yielding $\tauRX = 25$~ps, $\tauReh = 340$~ps, and $I_{\rm X}(0)/I_{\rm eh}(0) = 1.3$.
    The exciton (blue line) and \eh{} (dashed red line) components are also shown. 
    (b) Experimental dynamics of the optical orientation degree (dots) in a \MAFAPIxy{0.4}{0.6} crystal. The black line is a fit within the two-component model using Eq.~\eqref{SI:eq:P_oo_def} and the parameters obtained from panel (a). The exciton and \eh{} contributions are shown by the solid blue and dashed red lines, respectively. The following parameters are evaluated from the fit: $\PooX(0) = 0.60$, $\tauSX = 60$~ps, $\Pooeh(0) = 0.35$, and $\tauSeh = 450$~ps. Note that $\Pooeh(0) < \Poo(0) < \PooX(0)$, with the total initial optical orientation degree $\Poo(0)=0.55$.
        }
    \label{fig:model}
\end{figure*}

The time-integrated ptoluminescence (PL) spectra after pulsed excitation of the \MAFAPI{} samples measured at a temperature of $T = 1.6$\,K are shown in Figures~\ref{fig:TIPL}(a-c) by solid lines. The main photoluminescence peak is located at 1.465\,eV with a full width at half maximum (FWHM) of 6\,meV for $x = 0.1$, at 1.523\,eV with a FWHM of 8\,meV for $x = 0.4$, and at 1.629\,eV with a FWHM of 6\,meV for $x = 0.8$. At low temperatures, for lead halide perovskites the main line in time-integrated PL spectra commonly does not exhibit exciton characteristics, but instead originates from the recombination of spatially separated \eh{} pairs~\cite{kopteva_bayer2025PRB_MAPbI3_OO,kopteva_bayer2024advSci_FAPbI3_OO}. The exciton has a short recombination time and a comparatively small amplitude compared to the parameters of the long-lived recombination process of spatially separated \eh{} pairs. 

The exciton resonance can be identified in the time-resolved PL spectrum, measured right after pulsed excitation using a streak camera. The exciton resonance also manifests itself in the reflectivity spectrum. However, the small sample area and surface roughness complicate such measurements for the studied samples. The exciton photoluminescence after excitation by a laser pulse is shown by dashed lines in Figures~\ref{fig:TIPL}(a-c) for the different samples. The exciton resonance is located at $E_{\rm X} = 1.479$\,eV for $x = 0.1$, at $E_{\rm X} = 1.540$\,eV for $x = 0.4$, and at $E_{\rm X} = 1.638$\,eV for $x = 0.8$ (see Table~\ref{table:parameters_summary}). 

In Figure~\ref{fig:Eg_vs_x}, the exciton energies $E_{\rm X}$ at $T =1.6$\,K are plotted against the MA content of \MAFAPI{} crystals. The \MAPI{} single-crystal value is shown by the blue square~\cite{kopteva_bayer2025PRB_MAPbI3_OO}. Data on polycrystalline \MAPI{} and \FAPI{} from Refs.~\cite{galkowski_nicholas2016enenvsci_FAPI_MAPI_Eg_Ex,fang_anoniettaLoi2016lightSciAppl_FAPI_film_spectra} are also given for comparison, since no data on \FAPI{} single crystals at $T \approx 2$~K are available. The exciton resonance energy at $T=1.6$~K in Figure~\ref{fig:Eg_vs_x} follows a linear dependence on MA content, similarly to the room-temperature data reported in Ref.~\cite{weber_weller2016jmatchemA_FAMAPbI3_phase_transitions} (shown by the red line in Figure~\ref{fig:Eg_vs_x}), although the slopes differ significantly.

The exciton binding energies are similar in \FAPI{} and \MAPI{} ($14$~meV and $16$~meV, respectively)~\cite{galkowski_nicholas2016enenvsci_FAPI_MAPI_Eg_Ex}, hence the variation in the exciton energy in Figure~\ref{fig:Eg_vs_x} is provided almost entirely by the band gap shift. We evaluate band gap energies at $T= 1.6$\,K of $E_{\rm g} = 1.494$\,eV for $x = 0.1$, of $E_{\rm g} = 1.555$\,eV for $x = 0.4$, and of $E_{\rm g} = 1.653$\,eV for $x = 0.8$. At room temperature, the \MAFAPI{} band gap energy shifts linearly with increasing MA composition from $1.45$~eV ($x = 0$) up to $1.52 - 1.54$~eV ($x = 1$)~\cite{weber_weller2016jmatchemA_FAMAPbI3_phase_transitions}, see the red line in Figure~\ref{fig:Eg_vs_x}. Transfering these data to cryogenic temperatures is not straightforward because of the complex structural phase diagram of \MAFAPI{}~\cite{francisco-lopez_goni2020jphyschemC_FAMAPbI3_phase_diagram}.
Indeed, the phase transitions between the cubic, tetragonal, and orthorhombic crystal phases result in band gap shifts that can be as large as $0.1$~eV in \MAPI{} at $T=160$~K~\cite{kirstein_bayer2022acsphot_spin_w_nuclei_MAPbI3}.  

\begin{figure*}[t]
    \centering
    \includegraphics[]{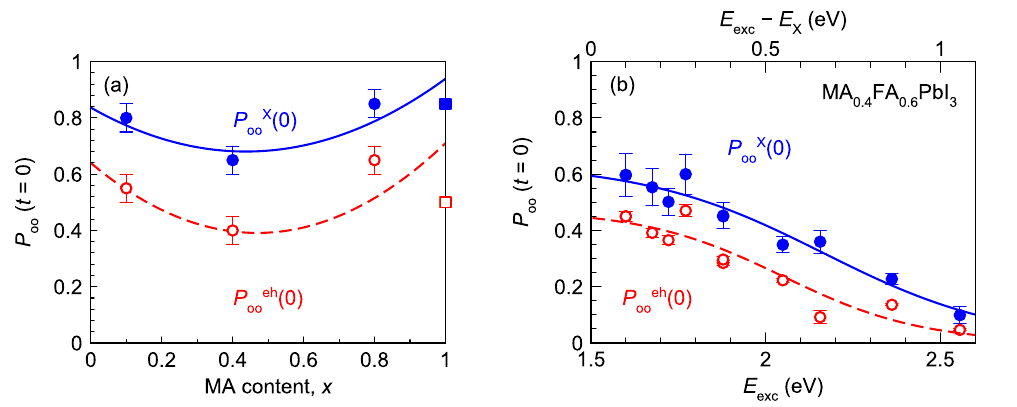}
    \caption{
    (a) Maximum exciton (closed blue symbols) and carrier (open red symbols) optical orientation degree as function of the MA content.
    The circles give the \MAFAPI{} data from this work, measured under nonresonant excitation with $P=0.5$~W/cm$^2$ at $T=1.6$~K, the squares give the data for \MAPI{} from Ref.~\cite{kopteva_bayer2025PRB_MAPbI3_OO}. 
    (b) Dependence of exciton and electron-hole optical orientation degree on excitation energy in a \MAFAPIxy{0.4}{0.6} crystal measured for $P = 0.5$~W/cm$^2$ at $T = 1.6$~K. The upper scale shows the laser detuning from the exciton resonance. In both panels, the lines are guides to the eye.}
    \label{fig:P_OO_model_P_OO_vs_x_P_OO_detuning}
\end{figure*}

\subsection{Recombination dynamics}
\label{sec:dynamics}

Time-resolved photoluminescence spectra are shown in Figure~\ref{fig:TRPL_x04}(a) for \MAFAPIxy{0.4}{0.6}, to give representative examples. Data for the other \MAFAPI{} samples are given in Figure~\ref{SI:fig:TRPL} of the Supporting Information (SI). The exciton emission is spectrally located in the range $1.536-1.545$~eV. For completeness, we also show the photoluminescence spectra taken in a 40~ps time window set to time delays of $t = 0$~ps, $100$~ps, $200$~ps, $500$~ps, and $12.5$~ns. Here, $t = 12.5$~ns corresponds to a negative delay of $t=-50$~ps, see Figure~\ref{fig:TRPL_x04}(b). 

In Figure~\ref{fig:TRPL_x04}(c), the PL dynamics averaged across three spectral ranges are shown.
In the exciton (X) range of $1.536-1.545$~eV, the dynamics follow a biexponential decay which are composed of the fast exciton decay with a decay time of $\tauRX=25$~ps followed by the extended recombination of \eh{} pairs with a decay time of $\tauReh=340$~ps. In the high energy (HE) range of $1.5405-1.545$~eV the dynamics are dominated by fast recombination with about 25~ps decay time, while in the low energy range of $1.525-1.536$~eV, where the PL intensity is strongest, the dynamics follow a long, nonexponential behavior, see the LL dynamics in Figure~\ref{fig:TRPL_x04}(c). This evidences that the PL in the LL range is dominantly contributed by the recombination of spatially-separated localized electrons and holes~\cite{kopteva_bayer2024advSci_FAPbI3_OO, kopteva_bayer2025PRB_MAPbI3_OO, kirstein_bayer2022natcomms_universal_g_perovskites}. 

The PL dynamics in the X ranges in the other \MAFAPI{} samples are shown in Figure~\ref{fig:TRPL_x04}(d). The exciton recombination on a time scale $\tauRX$ of about $25$~ps for all samples is accompanied by the longer decay of \eh{} pairs $\tauReh$ taking place over $200-400$~ps. The parameters of the recombination dynamics are summarized in Table~\ref{table:parameters_summary}. The coexistence of exciton  and \eh{} recombination is a common attribute of all the studied samples, as well as of \MAPI~\cite{kopteva_bayer2025PRB_MAPbI3_OO} and \CsTenFAPIBr{} crystals~\cite{kopteva_bayer2024advSci_FAPbI3_OO}. More details on the experimental features of the recombination dynamics for the samples with $x=0.1$ and $0.8$ are given in the SI, see Section~\ref{SI:trpl}.

\subsection{Optical Orientation}
\label{sec:results}

In optical orientation experiments, circularly polarized laser light is used to generate spin-oriented excitons and charge carriers. Measuring the optical orientation degree of the PL provides information on the spin initialization and relaxation. In the experiments, we use $\sigma^{+}$ circularly polarized laser pulses excitation and detect the dynamics of the PL intensity in either $\sigma^{+}$ or $\sigma^{-}$ polarizations.   
The degree of optical orientation is calculated by
\begin{align}
    \Poo(t) = \frac{I^{++}(t) - I^{+-}(t)}{I^{++}(t) + I^{+-}(t)}\,,
\label{eq:Poo}
\end{align}
where $I^{++}(t)$ and $I^{+-}(t)$ are the photoluminescence intensity dynamics measured in $\sigma^{+}$ and $\sigma^{-}$ polarizations, respectively. Spectrally-resolved dynamics of the optical orientation degree are shown in Figure~\ref{fig:TRPL_P_OO}(a). One can see that considerable optical orientation is detected only in the exciton spectral range of $1.536-1.545$~eV. More details are given in Figures~\ref{fig:TRPL_P_OO}(b,c), where the polarized PL spectra at zero delay time ($t=0$) and the resulting spectral dependence of $\Poo(t=0)$ are shown. 

The dynamics of the polarized PL components integrated over the exciton spectral range are shown in Figure~\ref{fig:TRPL_P_OO}(d). From these dynamics and Eq.~\eqref{eq:Poo} we calculate the time evolution of the optical orientation degree shown in Figure~\ref{fig:TRPL_P_OO}(e). Similarly to the recombination dynamics shown in Figure~\ref{fig:TRPL_x04}(d), the optical orientation dynamics has two components evidencing that two spin systems are contributing to the signal. Such behavior has also been observed in the other \MAFAPI{} samples, see Figure~\ref{SI:fig:TRPL_OO} in SI.
Indeed, it is typical also for bulk \MAPI{} and \CsTenFAPIBr{} lead halide perovskite crystals~\cite{kopteva_bayer2025PRB_MAPbI3_OO, kopteva_bayer2024advSci_FAPbI3_OO}. The two contributing spin systems can be identified as excitons and spatially separated \eh{} pairs. A biexponential fit used to estimate the exciton and \eh{}-pair initial optical orientation degrees gives $\PooX(0) = 0.50$ and $\Pooeh(0) = 0.35$ as well as their decay times of 15~ps and 460~ps.

\subsubsection{Optical orientation of excitons and carriers}

\label{sec:OO1}

When two spin systems contribute to the optical orientation dynamics in the same spectral range, the resulting spin dynamics can be nontrivial, as has been shown experimentally and theoretically in Ref.~\cite{kopteva_bayer2025PRB_MAPbI3_OO}. The dynamics are governed not only by the spin relaxation times, but also by the recombination times and the relative contribution of each system to the emission intensity. In particular, when the exciton recombination time $\tau_\mathrm{R}^\mathrm{X}$ is much shorter than the corresponding spin relaxation time $\tau_\mathrm{s}^\mathrm{X}$, the short component of the biexponential decay is governed by $\tau_\mathrm{R}^\mathrm{X}$ rather than $\tau_\mathrm{s}^\mathrm{X}$.

In the SI, Section~\ref{SI:model}, we present a detailed model of the optical orientation dynamics accounting for both the exciton and carrier contributions. The model allows us to distinguish between the exciton and the carrier contributions to the optical orientation dynamics and evaluate the respective initial polarizations $\PooX(0)$ and $\Pooeh(0)$ as well as the spin relaxation times $\tauSeh$ and $\tauSX$. Details of the model description of the \MAFAPIxy{0.4}{0.6} data are given in Figure~\ref{fig:model}. 

For the analysis, we start with a biexponential fit of the PL dynamics, in order to evaluate the recombination times and initial PL intensities. In Figure~\ref{fig:model}(a), the blue and red lines show the exciton and carrier dynamics, respectively, with $\tauRX = 25$~ps and $\tauReh = 340$~ps. The ratio of their initial PL intensities is $\initPLratio = 1.3$. 
These parameters are used for fitting the dynamics of the optical orientation degree with Eq.~\eqref{SI:eq:P_oo_def}. The resulting fit is shown in Figure~\ref{fig:model}(b) with the parameters for the excitons given by $\PooX(0) = 0.60$ and $\tauSX = 60$~ps, while for the carriers they are $\Pooeh(0) = 0.35$ and $\tauSeh = 450$~ps.

Comparing these parameters with the biexponential fit of the optical orientation dynamics in Figure~\ref{fig:TRPL_P_OO}(e), we find that the parameters characterizing the \eh{} pairs remain unchanged, because they are extracted from the long time component of the dynamics, when excitons have already recombined (compare $\Poo(t)$ and $\Pooeh(t)$ in Fig.~\ref{fig:model}(b)). In contrast, accounting for the initial exciton-to-carrier ratio $\initPLratio$ leads to a higher estimate of $\PooX(0)=0.60$ when Eq.~\eqref{SI:eq:P_oo_def} is used, compared with the 0.50 resulting from the biexponential fit in Figure~\ref{fig:TRPL_P_OO}(e).

Additionally, the evaluated $\tauSX = 60$~ps is significantly longer than the short $15$~ps time in the biexponential fit in Figure~\ref{fig:TRPL_P_OO}(e). The latter time is the result of the fast exciton recombination time and the exciton spin relaxation time which combine to $\left( 1/\tauRX + 1/\tauSX \right)^{-1} = 17$~ps. In the limit $\tauSX \to \infty$, the short time in the biexponential fit approaches $\tauRX$. We also note that here and throughout all the performed experiments, $\Pooeh(0) < \Poo(0) < \PooX(0)$.

Identical measurements and analyses are performed for all studied  \MAFAPI{} crystals with different compositions. Results on the maximum optical orientation degree $\Poo(t=0)$ for the excitons and the \eh{} pairs are plotted against MA content in Figure~\ref{fig:P_OO_model_P_OO_vs_x_P_OO_detuning}(a). Additionally, the data on \MAPI{} crystals from Ref.~\cite{kopteva_bayer2025PRB_MAPbI3_OO} are shown for comparison. In the \MAFAPI{} crystals, high optical orientation is present across different compositions, amounting to $60-80$\% for excitons and to $35-70$\% for carriers, see also Table~ \ref{table:parameters_summary}. The lowest optical orientation degree of 60\% for excitons and 35\% for carriers is found for $x=0.4$, where the effect of alloy fluctuations is expected to be maximal. 

\begin{figure*}[t]
    \centering
    \includegraphics[]{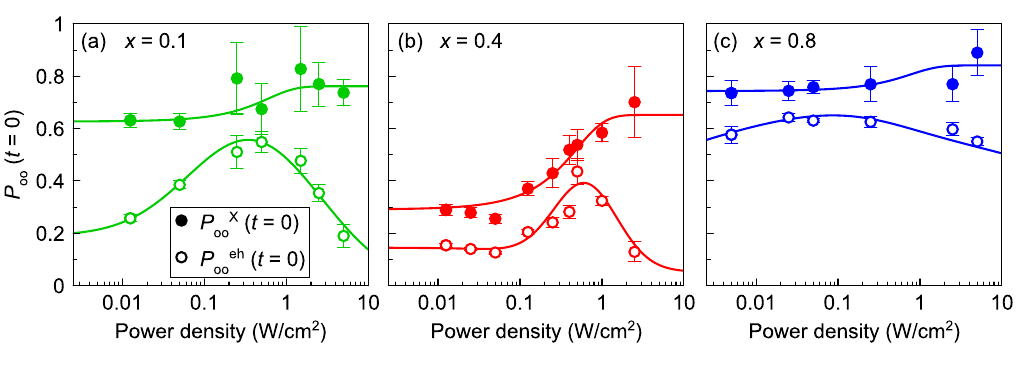}
    \caption{
    Dependence of initial exciton ($\PooX(0)$
    closed circles) and \eh{} ($\Pooeh(0)$, open circles) optical orientation degree on excitation density in \MAFAPI{} crystals with $x = 0.1$ (a), $x = 0.4$ (b), and $x = 0.8$ (c), measured using $E_{\rm exc} = 1.77$~eV excitation photon energy at $T=1.6$~K.
    Lines are guides to the eye.
    }
    \label{fig:P_OO_power}
\end{figure*}

\begin{figure*}[t]
    \centering
    \includegraphics[]{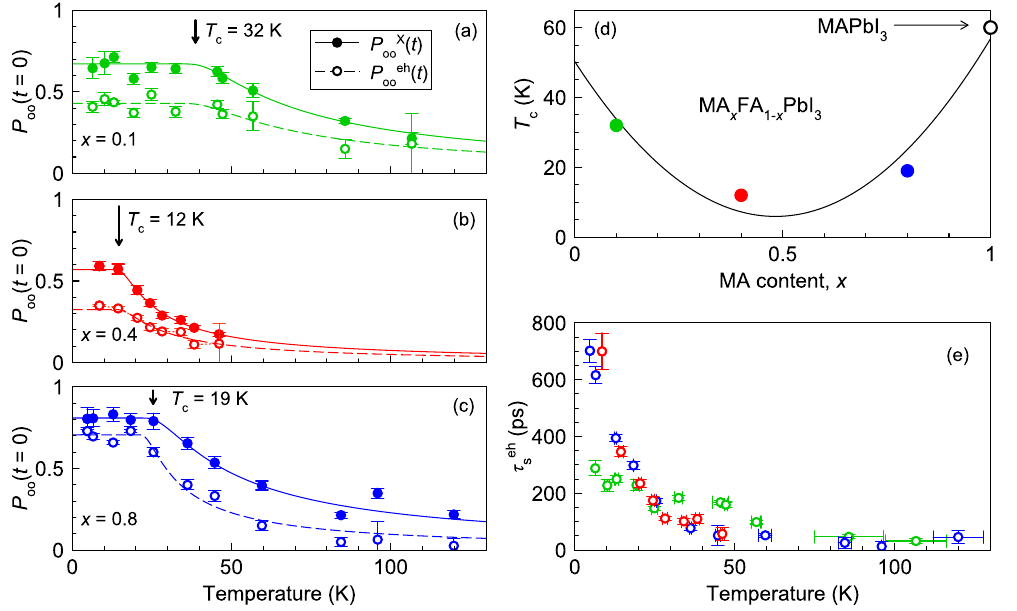}
    \caption{Temperature dependence of the optical orientation dynamics in \MAFAPI{} crystals. (a--c)~Optical orientation degree of excitons $\PooX(t=0)$ (closed circles) and carriers $\Pooeh(t=0)$ (open circles) in the samples with $x = 0.1$ for $E_\text{exc} = 1.550$~eV (a), $x=0.4$ for $E_\text{exc} = 1.560$~eV (b), and $x=0.8$ for $E_\text{exc} = 1.675$~eV (c) as function of temperature. The lines are guides to the eye. $P = 0.5$~\WcmSq{}. 
    (d)~Dependence of the critical temperature $T_{\rm c}$ on MA content. The closed circles are data from this work, the open circle is taken from Ref.~\cite{kopteva_bayer2025PRB_MAPbI3_OO}. The line is a guide for the eye.
    (e)~Temperature dependence of the electron-hole spin relaxation time.}
    \label{fig:P_OO_temperature}
\end{figure*}

For a closer look into the spin relaxation mechanisms responsible for the decrease of the optical orientation degree in \MAFAPIxy{0.4}{0.6}, we measure $\Poo(t=0)$ for different detunings of the excitation energy $E_{\rm exc}$ from the exciton resonance $E_{\rm X}$. The corresponding dependences for excitons and carriers are shown in Figure~\ref{fig:P_OO_model_P_OO_vs_x_P_OO_detuning}(b). Scanning the optical detuning uncovers the stability of the maximum value (about $60\%$) up to a detuning of about $0.3$~eV, highly reminiscent of what was observed for \CsTenFAPIBr{} and \MAPI{} crystals~\cite{kopteva_bayer2024advSci_FAPbI3_OO,kopteva_bayer2025PRB_MAPbI3_OO}. The results of Ref.~\cite{Kopteva_2025OOX} demonstrate that the behavior is expected for most bulk lead halide perovskites.

\subsubsection{Excitation power dependence of optical orientation}
\label{sec:power}

In order to study the possible effects of exciton-exciton and exciton-carrier scattering on the recombination and spin dynamics we measured the dynamics for various excitation densities in the range of $0.01-5$~\WcmSq{}. The results for time-resolved PL are given in the SI, Section~\ref{SI:power_dep} and Figure~\ref{SI:fig:MA0.4FA0.6PbI3_PL_vs_power}. The PL dynamics turn out to be similar in \MAFAPI{}{} crystals with different compositions.

The exciton recombination time $\tauRX$, which determines the initial decay, increases from about $10$~ps at low power to about $60$~ps at high power. This can be attributed to th exciton-exciton and exciton-carrier scattering at higher excitation densities. However, the $\tauReh$ time does not show a significant power dependence, remaining in the range of $200-400$~ps, although it somewhat differs from sample to sample. The intensity ratio $\initPLratio$ decreases with increasing power from about $7$ to about $1$.

The power dependences of $\PooX(0)$ and $\Pooeh(0)$, shown in Figure~\ref{fig:P_OO_power}, vary substantially across the studied \MAFAPI{} crystals. For excitons, the dependences are weak in the samples with $x=0.1$ and $0.8$, remaining in the range of $65-85$\% for all power density values. In contrast, for $x=0.4$, $\PooX(0)$ changes considerably, increasing from 30\% for low powers up to 65\% at $P > 1$~\WcmSq{}.
For carriers, the dependence of $\Pooeh(0)$ on excitation power is very weak in the \MAFAPIxy{0.8}{0.2} sample. However, it is strongly non-monotonic in the samples with $x=0.1$ and $0.4$, showing a decrease of the optical orientation degree at higher powers.

Despite the eventual drop in $\Pooeh(0)$ at high excitation power densities, the \eh{} spin relaxation time $\tauSeh$ is largely insensitive to the incident power in the covered range across all samples, see Figure~\ref{SI:fig:MA0.8FA0.2PbI3_OO_times_vs_power}(a) in the SI. On the other hand, $\tauSX$ tends to decrease from an effectively infinite value $\tauSX \gg \tauRX$ into the $100$~ps range, see Figure~\ref{SI:fig:MA0.8FA0.2PbI3_OO_times_vs_power}(b) in the SI.

Although the aforementioned two-component model generally predicts biexponential-like optical orientation dynamics, more complicated scenarios, including nonmonotonic behavior, are revealed in the case when exciton spin relaxation becomes comparable to exciton recombination time. This can occur when the exciton contribution to the PL dominates over that of the \eh{} pairs during a time $t \gtrsim \tauSX$ with $\PooX(t) < \Pooeh(t)$. A regime close to a non-monotonic optical orientation dynamics can be accessed, e.\,g., by subjecting the sample to sufficiently strong laser excitation. In our experiments, this behavior is most prominent for the \MAFAPIxy{0.8}{0.2} crystal. These peculiar optical orientation dynamics are shown in SI, see Figure~\ref{SI:fig:MA0.8FA0.2PbI3_OO_times_vs_power}(c) and Table~\ref{SI:table:parameters_three_powers}. The unusual shape of the dynamics permits a more accurate (to an error of $10\%$) measurement of $\tauSX$, as opposed to the more typical dynamics of Figure~\ref{fig:model}(b), which only permit an order-of-magnitude estimate at best. At $P \le 0.25$~\WcmSq{}, $\tauSX$ extends to over $200$~ps and exceeds $\tauRX=24$~ps by about an order of magnitude.

\begin{figure*}[t]
    \centering
    \includegraphics[]{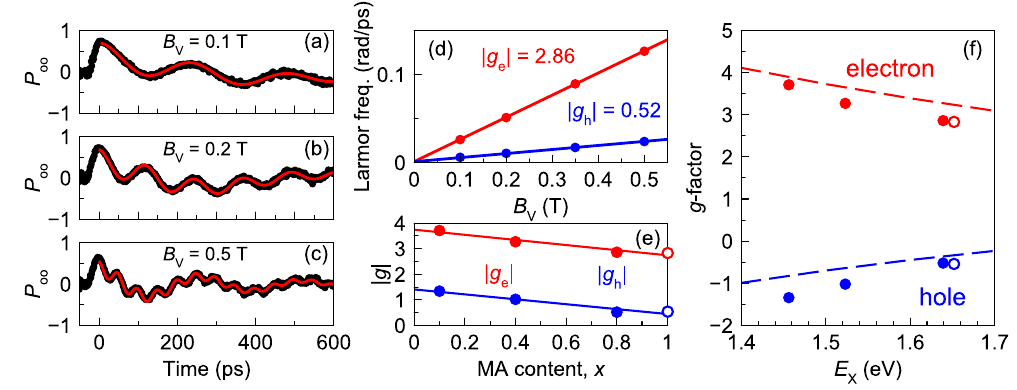}
    \caption{Optical spin orientation in Voigt magnetic field.
    (a--c) Optical orientation dynamics in a Voigt magnetic field of $B_{\rm V} = 0.1$\,T (a), $B_{\rm V} = 0.2$\,T (b), and $B_{\rm V} = 0.5$\,T (c), measured for the \MAFAPIxy{0.8}{0.2} sample at $T=1.6$~K (black dots). $E_{\rm exc} = 1.77$~eV, $P=0.5$~W/cm$^2$, and the detection energy is $1.638$~eV. The red lines are fits with Eq.~\eqref{eq:P_oo_BV_sum} with the parameters given in Table~\ref{table:parameters_Bv}.    
    (d) Magnetic field dependence of the Larmor precession frequencies of electrons and holes. The lines are linear fits yielding the absolute values of the electron and hole $g$-factors.
    (e) Dependence of the electron and hole $g$-factor values on the MA content.
    The $x=1$ data for \MAPI{} (open circles) are taken from Ref.~\cite{kopteva_bayer2025PRB_MAPbI3_OO}.
    (f) Electron and hole $g$-factors as function of the exciton energy $E_{\rm X}$. The closed circles are from this work, and the open circles are for \MAPI{} from  Refs.~\cite{kopteva_bayer2024advSci_FAPbI3_OO, kopteva_bayer2025PRB_MAPbI3_OO}.
    The dashed lines show the universal dependences of the electron and hole $g$-factors on band gap energy taken from  Ref.~\cite{kirstein_bayer2022natcomms_universal_g_perovskites}.
    Note that $E_{g}$ and $E_{\rm X}$ differ by the exciton binding energy (about 15~meV), which does not affect the dependence at this energy scale.
    }
    \label{fig:P_OO_beats}
\end{figure*}

\subsubsection{Temperature  dependence of optical orientation}
\label{sec:temperature}

The effect of temperature effect on the optical orientation degree and the spin dynamics is shown in Figure~\ref{fig:P_OO_temperature}. All samples demonstrate qualitatively similar temperature dependences of $\PooX(0)$ and $\Pooeh(0)$ [Figure~\ref{fig:P_OO_temperature}(a--c)]. The initial optical orientation degree of excitons and \eh{} pairs remains stable up to a certain sample-dependent temperature $T_{\rm c}$, beyond which it gradually decreases with the temperature increasing up to about $100$~K. The dependence of $T_{\rm c}$ on MA content is shown in Figure~\ref{fig:P_OO_temperature}(d), with the data on \MAPI{} crystals taken from Ref.~\cite{kopteva_bayer2025PRB_MAPbI3_OO}. A parabolic fit is shown as a guide to the eye. The lowest $T_{\rm c} = 12$~K is obtained for $x=0.4$, for which alloy fluctuations are expected to be the greatest. $T_{\rm c}$ is 32~K for $x=0.1$ and 19~K for $x=0.8$. 
Note, that in \MAPI{} crystals $T_{\rm c}$ reaches 60~K~\cite{kopteva_bayer2025PRB_MAPbI3_OO}, while in mixed alloy \CsTenFAPIBr{} crystals it is very low with $T_{\rm c} \approx 10$~K~\cite{kopteva_bayer2024advSci_FAPbI3_OO}.

The \eh{} spin relaxation time $\tauSeh$ shown in Figure~\ref{fig:P_OO_temperature}(e) shortens with temperature.
The exciton spin relaxation time $\tauSX$ cannot be reliably measured at these temperatures.
The exciton recombination time $\tauRX$ is not affected by temperature and remains about $10$~to $30$~ps for all samples, see Figure~\ref{SI:fig:MAFAPI_PL_vs_T}(a) in the SI. The \eh{} recombination time $\tauReh$ tends to increase approximately linearly with temperature, starting from about $300$~ps in all samples, see Figure~\ref{SI:fig:MAFAPI_PL_vs_T}(b) in the SI.
At about $100$~K, it reaches $500$~ps for $x = 0.1$ and $800$~ps for $x = 0.8$. In \MAFAPIxy{0.4}{0.6}, it increases up to $400-500$~ps at $50$~K. The ratio of exciton and carrier PL contributions $\initPLratio$ drops with temperature, as shown in Figure~\ref{SI:fig:MAFAPI_PL_vs_T}(c) in the SI. 

\subsection{Spin beats in Voigt geometry}
\label{sec:beats}

Figure~\ref{fig:P_OO_beats}(a) presents the temporal evolution of $\Poo(t)$ detected at the exciton energy of MA$_{0.8}$FA$_{0.2}$PbI$_{3}$ in a magnetic field of $B_\mathrm{V}=0.1$~T applied perpendicular to the light wave vector (Voigt geometry). The signal exhibits a complex spin-beat pattern with a slow decay, extending over 600~ps. This decay time substantially exceeds the exciton lifetime, indicating that the observed dynamics originate from the coherent spin precession of spatially separated localized electrons and holes. The dynamics of the optical orientation degree measured for \MAFAPIxy{0.8}{0.2} at $B_{\rm V} = 0.2$\,T and $0.5$\,T are shown in Figures~\ref{fig:P_OO_beats}(b,c). They have two components with Larmor precession frequencies corresponding to electron and hole spin precession. The red lines are fits with~\cite{kochereshko_lavallard1998physsolidstate_Pe_Ph_beats}
\begin{align}
    \Pooeh(t) = \frac{\Poo^{\rm e}(t) + \Poo^{\rm h}(t)}{1 + \Poo^{\rm e}(t) \Poo^{\rm h}(t)},
\label{eq:P_oo_BV_sum}
\end{align}
with
\begin{align}
\Poo^{\rm e(h)}(t) = \Poo^{\rm e(h)}(0) \cos [\omega_{\rm e(h)} t] \exp [-t/T_{2}^{\rm *,e(h)}],
\label{eq:P_oo_BV_comp}
\end{align}
where $\Poo^{\rm e(h)}(0)$ is the initial electron (hole) optical orientation degree, $\omega_{\rm e(h)}$ is the Larmor precession frequency of electron (hole), and $T_{2}^{\rm *,e(h)}$ is the spin dephasing time of electron (hole). The parameters used to fit the optical orientation dynamics in Figures~\ref{fig:P_OO_beats}(a--c) are given in Table~\ref{table:parameters_Bv}. The fits provide an estimate of the $T_{2}^{\rm *,e(h)}$ times of the order of $500$~ps at $B_{\rm V} = 0.1$~T for all samples.

\begin{table}[htbp]
\renewcommand{\arraystretch}{1.5}
\caption{Parameters used to fit the dynamics of the optical orientation degree shown in Figures~\ref{fig:P_OO_beats} (a--c). }
\begin{ruledtabular}
\begin{tabular}{ccccccc}
$B_{\rm V}$ (T) & \multicolumn{2}{c}{$\Poo(0)$} 
                & \multicolumn{2}{c}{$\omega$ (rad/ps)} 
                & \multicolumn{2}{c}{$T_{2}^{*}$ (ps)} \\ \hline
    & $e$ & $h$ & $e$ & $h$ & $e$ & $h$ \\
0.1 & 0.36 & 0.37 & 0.026 & 0.006 & 500 & 500 \\
0.2 & 0.32 & 0.40 & 0.051 & 0.011 & 460 & 410 \\
0.5 & 0.18 & 0.40 & 0.126 & 0.024 & 360 & 380 \\
\end{tabular}
\end{ruledtabular}
\label{table:parameters_Bv}
\end{table}

The magnetic field dependences of the Larmor precession frequencies of electrons and holes are linear without a zero-field offset, see Figure~\ref{fig:P_OO_beats}(d). Fitting them with $\hbar \omega_{\rm e(h)} = |g_{\rm e(h)}| \mu_{\rm B} B_{\rm V}$  gives the values of electron and hole $g$-factors in \MAFAPIxy{0.8}{0.2}, $|g_{\rm e}|=2.86$ and $|g_{\rm h}|=0.52$. $\mu_{\rm B}$ is the Bohr magneton. The $g$-factors for all \MAFAPI{} crystals are given in Table~\ref{table:parameters_summary} and are plotted as a function of MA content in Figure~\ref{fig:P_OO_beats}(e). These dependences are approximately linear with $x$, showing that Vegard's law holds for the electron and hole $g$-factors in \MAFAPI{} crystals.

It has been found in previous experimental and theoretical studies~\cite{kirstein_bayer2022natcomms_universal_g_perovskites} that in lead halide perovskite semiconductors with a band gap smaller than $1.82$~eV,  $g_{\rm e}>0$ and $g_{\rm h}<0$. In Figure~\ref{fig:P_OO_beats}(f) we plot the measured $g$-factors with account for their signs as function of the exciton energy $E_{\rm X}$, which serves as an estimate of the band gap $E_{g}$. Data for \MAPI{} crystals are also shown by the open circles. The dashed lines show the universal dependence of the charge carrier $g$-factors on the band gap energy, as established in Ref.~\cite{kirstein_bayer2022natcomms_universal_g_perovskites}. One can see that the mixed-cation \MAFAPI{} crystals follow to a good approximation this universal dependence.

\subsection{Discussion}
\label{sec:Discusion}

In our experiments, the optical orientation of excitons and charge carriers is measured under nonresonant excitation with substantial detuning from the exciton resonance. Under these conditions, both energy and spin relaxation proceed through several processes within the lifetime. Specifically, spin relaxation may occur at three stages: (i) during the initial energy relaxation mediated by LO phonon emission, (ii) during the subsequent energy relaxation near the band edge via acoustic phonon scattering, and (iii) through spin relaxation of thermalized excitons and carriers, which are localized in the vicinity of the band extrema.

The initial fast energy relaxation is assisted by LO phonon emission. Our results presented in Figure~\ref{fig:P_OO_model_P_OO_vs_x_P_OO_detuning}(b) demonstrate that the spin relaxation during this stage is very weak. The dependences of the optical orientation degree on the laser energy detuning are qualitatively similar for excitons and charge carriers. In line with our conclusions in Refs.~\cite{kopteva_bayer2024advSci_FAPbI3_OO,Kopteva_2025OOX}, such detuning dependences provide strong evidence for the absence of the Dyakonov–Perel spin relaxation mechanism and, therefore, for the conservation of spatial inversion symmetry in \MAFAPI{} crystals.

The further energy relaxation in the vicinity of the band gap, preceding capture into localized states, is assisted by acoustic phonons. At this stage, scattering of free excitons and carriers by alloy fluctuations may accelerate energy relaxation by opening additional channels for momentum relaxation. Consequently, the characteristics of the fluctuation potential, which are also determined by the composition parameter $x$, become important. However, these characteristics are not yet well understood and require detailed investigations. Scattering on acoustic phonons gives rise to Elliott–Yafet spin relaxation.

A specific feature of the lead halide perovskite semiconductors is the presence of long-lived resident carriers, which may originate from photogeneration. At cryogenic temperatures, resident electrons and holes are localized at separate locations. Free excitons and charge carriers can scatter on the resident carriers, and exchange with them their spin polarization. Then the spin polarization of the localized resident carriers can relax via their interaction with nuclear spin system. For free photogenerated carriers this process may provide spin relaxation which is analogous to the Bir–Aronov–Pikus mechanism.
This process is expected to be most efficient for $x \approx 0.5$, where the fluctuation potential is anticipated to have the largest amplitude as well as the highest density of localized states. Consequently, a higher concentration of resident carriers can be expected. This interpretation is consistent with our experimental data, where the minimum in the optical orientation degree is observed for both excitons and carriers in the $x = 0.4$ sample, see  Figure~\ref{fig:P_OO_model_P_OO_vs_x_P_OO_detuning}(a).

For localized carriers, hyperfine interaction with the nuclear spin system leads to the relaxation of localized carrier spin orientation. In lead halide perovskite semiconductors this mechanism has been investigated experimentally for \CsTenFAPIBr{} crystals and analyzed theoretically in Ref.~\onlinecite{kudlacik2024optical}. In the absence of an external magnetic field, the electron and hole spins precess about the Overhauser field of the nuclear spin fluctuations within the carrier localization volume. As a result, the component of the carrier spin polarization perpendicular to the direction of this Overhauser field relaxes with a characteristic time proportional to $\omega_{\rm e(h),r}^{-1}$, where $\omega_{\rm e(h),r}$ is the Larmor precession frequency of the electron (hole) spin precession in the random Overhauser field. For \CsTenFAPIBr{} crystals, the corresponding spin relaxation times are of the order of 100~ps for electrons and holes~\cite{kudlacik2024optical}. Further decay is slow, and is controlled by the correlation time of the carrier-nuclear spin system. These timescales are therefore considerably longer than those of the carrier cooling and localization processes, which are completed within $5-10$~ps. 

In view of the mechanisms discussed above, we propose that the initial degree of optical orientation, $\Poo(0)$, is mainly determined by spin relaxation processes occurring during the energy relaxation assisted by acoustic phonons, i.e., prior to carrier localization. The smaller value of $\Poo(0)$ observed for the $x=0.4$ sample can be reasonably attributed to the enhanced contribution of alloy fluctuations to carrier scattering. At longer times, the carrier spin relaxation dynamics, characterized by the time constant $\tauSeh$ of about $400-600$~ps, is likely governed by spin relaxation of localized electrons and holes in the random Overhauser field created by nuclear spin fluctuations.  

\section{Conclusions}
\label{sec:conclusions}

On the basis of the presented experimental results and their discussion, we conclude that the spin-dependent properties of \MAFAPI{} crystals are considerably influenced by the relative fraction of MA and FA. We suggest that for the available set of data, the main effect of the mixed-cation nature of the samples comes at the stage of exciton and carrier energy relaxation via acoustic phonons in the vicinity of the band gap.

For clarifying further details and for specific conclusions about the role of alloy fluctuations on exciton/carrier localization and on spin dynamics, more experimental and theoretical studies are needed. Here, the involvement of other experimental techniques that are by now established in the spin physics of lead halide perovskites are required. Among them are time-resolved Faraday/Kerr rotation~\cite{Kirstein_FAPI_2022}, spin-flip Raman scattering~\cite{kirstein_bayer2022natcomms_universal_g_perovskites}, optical orientation under continuous wave excitation for highlighting the effects of the carrier-nuclei interaction~\cite{kudlacik2024optical}, spin-dependent photon echo~\cite{Grisard_2024FAPI}. Additionally, a broader spectrum of materials needs to be studied, e.g., mixed-cation materials with substitutions of MA-Cs and FA-Cs and mixed-halogen materials like MAPb(I,Br)$_3$ or CsPb(Br,Cl)$_3$.

\section{Experimental Section}
\label{sec:experimental}

{\it Samples:} The crystals under study are grown from solution by the well-established inverse crystallization method~\cite{chen_bakr2019acsenlett_ITC_MAPbI3_PV, alsalloum_bakr2020acsenlett_ITC_MAPbI3_PV}. This method is known to produce high-quality single crystals with well-defined facets and long carrier diffusion lengths~\cite{turedi_bakr2022advmat_ITC_PV_long_diff_lengths}. X-ray diffraction measurements show that these perovskite single crystals have high structural quality~\cite{yang_mohammed2022acsenlett_ITC_xray_hiquality}. The studied \MAFAPI{} single crystals with $x = 0.1$, $0.4$, and $0.8$ were synthesized from appropriately mixed MAI, FAI, and PbI$_2$ perovskite precursors. The precursors were injected between two polytetrafluoroethylene coated glasses and slowly heated to 120$^{\circ}$C. The samples studied in this work have square shapes tallying up to about $2 \times 2$~mm in the (001) crystallographic plane and a thickness of about 30~$\mu$m.

{\it Time-resolved photoluminescence and optical orientation:}
The samples are excited by a mode-locked laser (Chameleon Discovery, tunable spectral  range of $E_{\rm exc} = 1.30-2.55$~eV, repetition rate of $80$~MHz, pulse duration of $100$~fs), whose emission is sent through a  Glan-Taylor  polarizer and a quarter-wave plate to ensure $\sigma^{+}$ polarization of excitation. The excitation spot on the samples has an area of $2\times 10^{-3}$~cm$^{2}$ ($500$~$\mu$m diameter). The incident laser power is measured before the focusing lens and was typically about $1$~mW, corresponding to a fluence of $P=0.5$~\WcmSq{}, which is small enough to exclude any nonlinear optical effects. The samples, held in strain-free paper envelopes, are kept in a liquid helium cryostat with a variable temperature insert ($T = 1.6-300$~K). At $T = 1.6$~K, the samples are immersed in superfluid helium, while at $T > 4$~K they are kept in helium vapor. The photoluminescence is collected through a lens, and the $\sigma^{+}$/$\sigma^{-}$ components are analyzed before being sent to an 0.5-meter Acton spectrometer. It is recorded either time-integrated using a charge coupled device (CCD) camera, or time-resolved using a Hamamatsu streak camera. A $300$~lines/mm grating was used for all measurements, as it provides an optimal compromise between spectral resolution and temporal resolution which is about 6 ps, as estimated by the PL rise time.

{\it Magneto-optical measurements:} A superconducting split-coil magnet generates magnetic fields up to $B_{\rm V} = 7$~T in Voigt geometry, i.\,e. $B_{\rm V}$ is orthogonal to the wave vector of the collected photoluminescence.


\section*{Acknowledgements}
The authors are thankful to M. M. Glazov and K. V. Kavokin for fruitful discussions. N.E.K. acknowledges the support of the Deutsche Forschungsgemeinschaft (project KO 7298/1-1, no. 552699366). I.A.A. acknowledges the support of the Deutsche Forschungsgemeinschaft (AK 40/13-1, no. 506623857). I.V.K. acknowledges Deutsche Forschungsgemeinschaft via project (KA 6253/1-1, no. 534406322). The work at ETH Z\"urich (B.T. and M.V.K.) was financially supported by the Swiss National Science Foundation (grant agreement 200020E 217589) through the DFG-SNSF bilateral program and by ETH Z\"urich through ETH+ Project SynMatLab.

\section*{Conflict of Interest}
The authors declare no conflict of interest.

\section*{Data Availability Statement}
The data that support the findings of this study are available from the corresponding author upon reasonable request.

\section*{Keywords}
perovskite semiconductors, mixed-cation perovskite crystals, optical spin orientation, spin dynamics of excitons, spin dynamics of charge carriers, time-resolved photoluminescence

%


\clearpage

\onecolumngrid
\setcounter{page}{1}
\setcounter{section}{0}
\setcounter{figure}{0}
\setcounter{table}{0}
\setcounter{equation}{0}
\renewcommand{\thesection}{S\arabic{section}}
\renewcommand{\thepage}{S\arabic{page}}
\renewcommand{\thefigure}{S\arabic{figure}}
\renewcommand{\thetable}{S\arabic{table}}
\renewcommand{\theequation}{S\arabic{equation}}

\begin{center}
\textbf{\large Supporting Information:}\\[4pt]
\textbf{\large Spin dynamics of excitons and carriers in mixed-cation \MAFAPI{} perovskite crystals: alloy fluctuations probed by optical orientation}\\[6pt]
B. F. Gribakin, N. E. Kopteva, D. R. Yakovlev, I. A. Akimov, I. V. Kalitukha, B.~Turedi,\\  M.~V.~Kovalenko, and M. Bayer
\end{center}

\section{Time-resolved photoluminescence and optical orientation dynamics} 
\label{SI:trpl}

\begin{figure*}[b]
    \centering
    \includegraphics[]{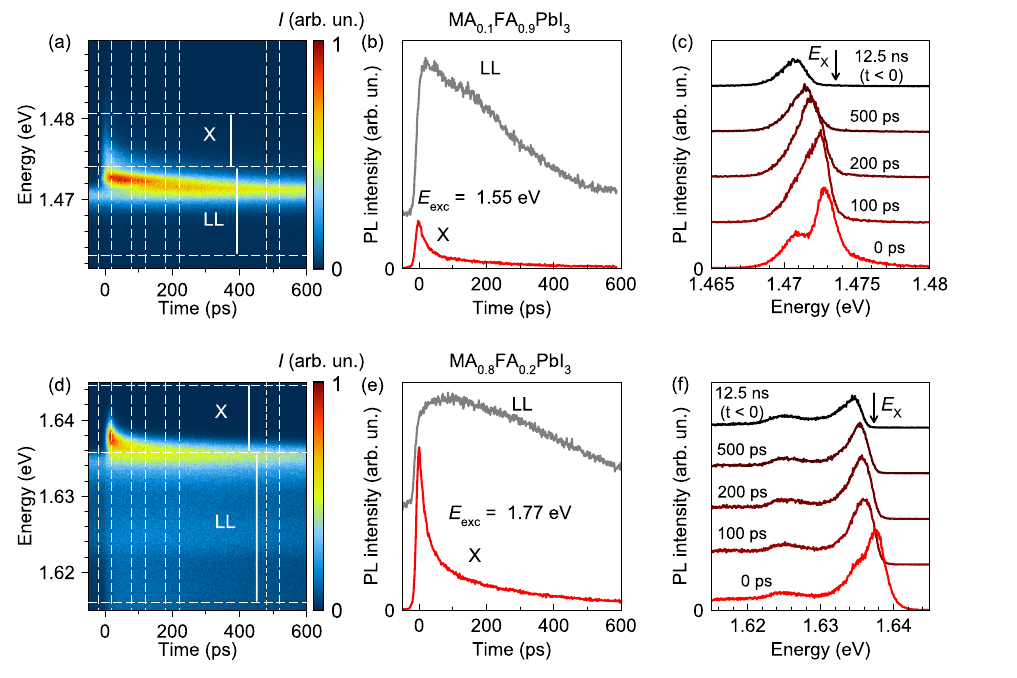}
    \caption{Time-resolved PL dynamics of \MAFAPIxy{0.1}{0.9} [panels (a--c)] and \MAFAPIxy{0.8}{0.2} [panels (d--f)] crystals measured at $E_\text{exc}=1.55$~eV and $1.77$~eV, respectively, using $P = 0.5$~\WcmSq{} at $T = 1.6$~K.
    (a,d) Color-coded plots of time-resolved PL. The red color corresponds to strong intensity. The horizontal dashed lines mark the high- and low-energy boundaries of the spectral ranges LL and X (shown by solid lines) used for the time-resolved data in panels (b,e), respectively. The vertical dashed lines mark the temporal ranges used for the time-integrated data presented in panels (c,f), respectively.
    (b,e) Spectrally-integrated PL dynamics in the X spectral range, see the ranges marked in panels (a,d).
    (c,f) PL spectra measured in a 40~ps time window at different delay times as marked by the dashed lines in panels (a,d). The PL spectra are shifted vertically for clarity.
    }
    \label{SI:fig:TRPL}
\end{figure*}

\begin{figure*}[ht]
    \centering
    \includegraphics[]{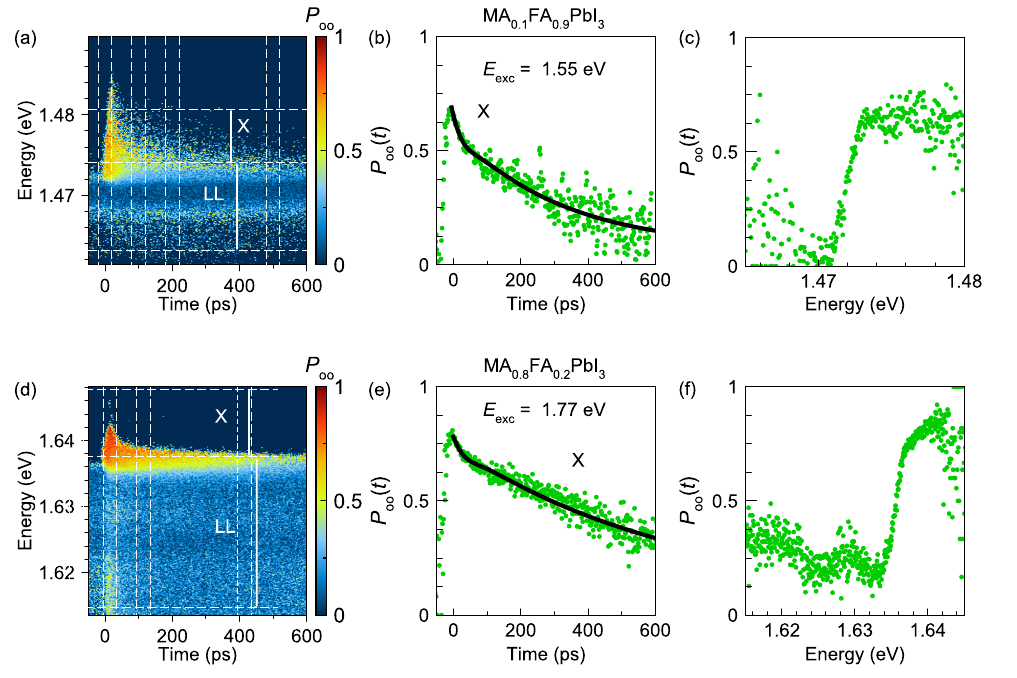}
    \caption{Time-resolved optical orientation dynamics of \MAFAPIxy{0.1}{0.9} [panels (a,b)] and \MAFAPIxy{0.8}{0.2} [panels (d,e)] crystals measured at $E_\text{exc}=1.55$~eV and $1.77$~eV, respectively, using $P = 0.5$~\WcmSq{} at $T = 1.6$~K.
    (a,d) Color-coded plots of time-resolved optical orientation. Red color corresponds to high polarization degree. 
    The horizontal dashed lines mark the high- and low-energy boundaries of the spectral ranges  LL and X (shown by solid lines) used for the time-resolved data in panels (b,e), respectively.
    The vertical dashed lines mark the temporal ranges used for the time-integrated data presented in panels (c,f), respectively.   
    (b,e) Spectrally-integrated optical orientation dynamics in the X spectral region [see ranges marked in panels (a,d)] (green circles) fitted by the two-component model (black lines). For the fit parameters, see Table~\ref{SI:table:parameters_TRPL_OO}.
    (c,f) Optical orientation spectra measured in a 40~ps time window at $t = 0$~ps as marked by the dashed lines in panels (a,d).
    }
    \label{SI:fig:TRPL_OO}
\end{figure*}

The time-resolved PL measured on the \MAFAPIxy{0.1}{0.9} and \MAFAPIxy{0.8}{0.2} samples is presented in Figure~\ref{SI:fig:TRPL}.
The data from Figures~\ref{SI:fig:TRPL}(a,d) are spectrally and temporally averaged, with the results shown in Figures~\ref{SI:fig:TRPL}(b,e) and Figures~\ref{SI:fig:TRPL}(c,f), respectively. The samples are excited using $\sigma^{+}$ polarized $100$~fs pulses.
The $\sigma^{+}$ and $\sigma^{-}$ components of the PL are recorded separately, then summed in order to exclude any artifacts and focus on the recombination dynamics.

Overall, the behavior is qualitatively similar for all studied samples including \MAFAPIxy{0.4}{0.6} (see Figure~\ref{fig:TRPL_x04}), with the signal clearly separable into a short-lived component and a long-lived component with spectral overlap between the two.
In Figure~\ref{SI:fig:TRPL}(a), the $x=0.1$ sample shows an apparent time-dependent shift of the main spectral peak that is not found in the other studied samples. This shift is present for all the applied laser fluences from $0.01$~\WcmSq{} to  $5$~\WcmSq{}, but disappears with temperature increase as all features in the PL spectrum broaden. The $x=0.8$ sample has a distinct trail of weak low-energy PL that seems to contain both short and long-lived components, see Figure~\ref{SI:fig:TRPL}(d), but it does not seem to affect the dynamics of the X spectral range in \MAFAPIxy{0.8}{0.2}.

The optical orientation dynamics extracted from the same experiments as shown in Figure~\ref{SI:fig:TRPL} are presented in Figure~\ref{SI:fig:TRPL_OO}. Here, the data are calculated from the $\sigma^{+}$ and $\sigma^{-}$ PL components using Eq.~\eqref{eq:Poo}. The results are presented in Figures~\ref{SI:fig:TRPL_OO}(a,d) as color-coded plots.
Both samples show a high ($>70\%$) initial polarization in the X spectral region which then decays about biexponentially [see Figures~\ref{SI:fig:TRPL_OO}(b,e)], as described in Section~\ref{sec:OO1} of the main text.
The spectrally averaged optical orientation dynamics shown in Figures~\ref{SI:fig:TRPL_OO}(b,e) are fitted by the two-component model with the fit parameters listed in Table~\ref{SI:table:parameters_TRPL_OO}. Spectrally, the short-lived emission is highly polarized at $t=0$~ps, while the long-lived emission is typically only weakly polarized, as shown in Figures~\ref{SI:fig:TRPL_OO}(c,f).

\begin{table}[t]
\renewcommand{\arraystretch}{1.5}
\caption{Parameters used to fit the dynamics of the optical orientation degree measured on \MAFAPI{} crystals at $T=1.6$~K using $P = 0.5$~\WcmSq{}. 
All fit parameters except $\tauSX$ are well-constrained with a precision of $10\%$ or better.
The estimated errors for $\tauSX$ are given in the table. The fits are shown by the solid lines in Figures~\ref{SI:fig:TRPL_OO}(b,e) and Figure~\ref{fig:model}(b).
}
\begin{ruledtabular}
\begin{tabular}{cccccccc}
Sample & \multicolumn{2}{c}{$\Poo(0)$}
                 & \multicolumn{2}{c}{$\tau_{\rm s}$ (ps)}
                 & \multicolumn{2}{c}{$\tau_{\rm R}$ (ps)} 
                 & $\initPLratio$ \\ \hline
     &  X   & $eh$ &  X  & $eh$& X  & $eh$& \\
\MAFAPIxy{0.1}{0.9} & 0.67 & 0.54 & $120 \pm 70$ & 320 & 22 & 230 & 2.6 \\
\MAFAPIxy{0.4}{0.6} & 0.60 & 0.35 & $60 \pm 30$ & 450 & 25 & 340 & 1.3 \\
\MAFAPIxy{0.8}{0.2} & 0.74 & 0.68 & $170 \pm 60$ & 650 & 23 & 320 & 2.5 \\
\end{tabular}
\end{ruledtabular}
\label{SI:table:parameters_TRPL_OO}
\end{table}

\section{Modeling of optical orientation dynamics contributed by excitons and carriers} 
\label{SI:model}

The spin dynamics of the excitons and charge carriers in semiconductors are controlled by their spin lifetime $T_{\rm s}$, which is determined by the recombination time $\tau_{\rm R}$ and the spin relaxation time $\tau_{\rm s}$: 
\begin{align}
\frac{1}{T_{\rm s}}=\frac{1}{\tau_{\rm R}} + \frac{1}{\tau_{\rm s}}. 
\end{align}
In case when only one system contributes to the spin dynamics, the dynamics of the optical orientation degree 
\begin{align}
\Poo(t)=\Poo(0) \exp(-t/\tau_{\rm s}) 
\end{align}
is controlled by the spin relaxation time only, which then can be directly evaluated. Note that the recombination time is not contributing here, as according to Eq.~\eqref{eq:Poo} it drops out. 

For the case of two spin systems contributing to the spin dynamics at the same spectral energy, the $\Poo(t)$ dynamics are contributed by the recombination times and the relative PL intensities of the two systems. They can have two exponentially decaying components, or even become non-monotonic. We have considered this case in Ref.~\cite{kopteva_bayer2025PRB_MAPbI3_OO} and give here more details, including the model descriptions of various regimes.

\begin{figure*}[t]
    \centering
    \includegraphics[]{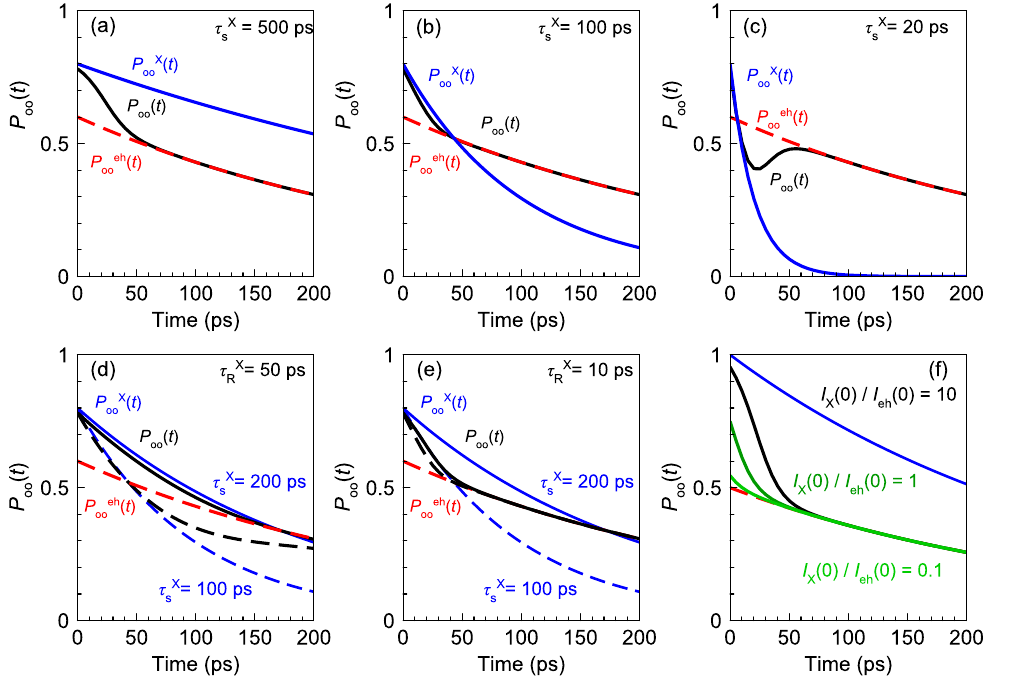}
    \caption{
    Different kind of behaviors of optical orientation as resulting from the phenomenological two-component model.
    In all panels, the exciton optical orientation $\PooX$ dynamics are shown by the blue solid lines, the \eh{}-pair $\Pooeh$ dynamics are shown by the red dashed lines, and the averaged $\Poo$ dynamics are shown by the solid black lines [light and dark green in panel (f)].
    (a,b,c) $\PooX(t)$ and $\Poo(t)$ are calculated for three values of the exciton spin relaxation time $\tauSX$, while the other parameters are fixed at $\tauRX = 10$~ps, $\tauReh = 1000$~ps, $\tauSeh = 300$~ps, $\PooX(0) = 0.8$, $\Pooeh(0) = 0.6$, and $I_{\rm X}(0) / I_{\rm eh}(0) = 10$.
    (d,e) The exciton spin relaxation time $\tauSX$ is set to $100$~ps or $200$~ps (solid or dashed lines, respectively), and the exciton recombination time $\tauRX$ is either $50$~ps [panel (d)] 
    or $10$~ps [panel (e)]; the other parameters are as in (a--c).
    (f) The relative number of excitons and \eh{} pairs is varied, keeping $\tauRX = 10$~ps, $\tauSX = 300$~ps, $\tauReh = 1000$~ps, $\tauSeh = 300$~ps, $\PooX(0) = 1$, and $\Pooeh(0) = 0.5$.
    The ratio $I_{\rm X}(t) / I_{\rm eh}(t)$ is time-dependent and approaches $0$ at $t \gg \tauRX$ as $\tauRX \ll \tauReh$.}
    \label{SI:fig:P_OO_model}
\end{figure*}

We consider in the model two spin systems (the systems of the excitons and spatially separated \eh{} pairs) which emission overlap spectrally. By modeling the experimental dynamics of $\Poo(t)$, this allows us to evaluate the initial spin polarizations of both systems ($\PooX(0)$ and $\Pooeh(0)$) and the spin relaxation time of the \eh{} pairs $\tauSeh$, as well as to estimate the exciton spin relaxation time $\tauSX$. In contrast to single-component optical orientation, where only a monoexponential decay is expected, here various shapes of optical orientation dynamics can be expected, from about biexponential to non-monotonic. This atypical behavior is controlled by the sensitive interplay of the model parameters, and can, for instance, permit the evaluation of the $\tauSX$ times even if $\tauRX \ll \tauSX$ in the case of a non-monotonic optical orientation dynamics.



If the excitation is $\sigma^{+}$ polarized, then the PL is expected to be preferentially $\sigma^{+}$ polarized due to optical orientation~\cite{opticalOrientation1984}. When excitons (\eh{} pairs) recombine, their emission intensity is $I_{\rm X}^{+}$ ($I_{\rm eh}^{+}$) in $\sigma^{+}$ polarization and $I_{\rm X}^{-}$ ($I_{\rm eh}^{-}$) in $\sigma^{-}$ polarization.
Note that these intensities are time-dependent, but we omit showing explicitly the time dependence for brevity.
For excitons, the degree of optical orientation is given by
\begin{align} \label{SI:eq:P_oo^X_def}
    \PooX (t) = \frac{I_{\rm X}^{+} - I_{\rm X}^{-}}{I_{\rm X}^{+} + I_{\rm X}^{-}} = \frac{I_{\rm X}^{+} - I_{\rm X}^{-}}{I_{\rm X}},
\end{align}
where $I_{\rm X}^{\phantom{+}} = I_{\rm X}^{+} + I_{\rm X}^{-}$. Similarly, for the circular polarization degree of the \eh{} pairs
\begin{align} \label{SI:eq:P_oo^eh_def}
    \Pooeh(t) = \frac{I_{\rm eh}^{+} - I_{\rm eh}^{-}}{I_{\rm eh}^{+} + I_{\rm eh}^{-}} = \frac{I_{\rm eh}^{+} - I_{\rm eh}^{-}}{I_{\rm eh}},
\end{align}
where $I_{\rm eh}^{\phantom{+}} = I_{\rm eh}^{+} + I_{\rm eh}^{-}$.

The total, or average, optical orientation degree accounting for both systems is given by 
\begin{align}
    \Poo(t) =& \frac{I_{\rm X}^{+} - I_{\rm X}^{-} + I_{\rm eh}^{+} - I_{\rm eh}^{-}}{I_{\rm X} + I_{\rm eh}} \nonumber\\
    &= \PooX \frac{I_{\rm X}(t)}{I_{\rm tot}(t)} + \Pooeh \left[1 - \frac{I_{\rm X}(t)}{I_{\rm tot}(t)} \right], \label{SI:eq:P_oo_def}
\end{align}
where $I_{\rm tot}(t) = I_{\rm X}(t) + I_{\rm  eh}(t)$.
We stress that the exciton-to-carrier intensity ratio $I_{\rm X}(t) / I_{\rm eh}(t)$ is time dependent.
In fact, this significantly impacts the optical orientation dynamics, as even a high initial ratio $\initPLratio$ gradually approaches zero at longer delays because the exciton recombination time is an order of magnitude shorter than that of the \eh{} pairs.

The excitons and \eh{} pairs are characterizeded by their respective spin relaxation times $\tauSX$ and $\tauSeh$, and recombination times $\tauRX$ and $\tauReh$. These times govern the decay of the optical orientation and population for the individual systems of excitons and \eh{} pairs.
\begin{align}
    \PooX(t) &= \PooX(0) \exp (-t / \tauSX), \\
    I_{\rm X}(t) &= I_{\rm X}(0) \exp (-t / \tauRX), \\
     \Pooeh(t) &= \Pooeh(0) \exp (-t / \tauSeh), \\
    I_{\rm eh}(t) &= I_{\rm eh}(0) \exp (-t / \tauReh), 
\end{align}

The typical recombination times at $T = 2$~K, as evaluated from biexponential fits of the PL dynamics and as reported in the main text [also see Table~\ref{table:parameters_summary} and Figures~\ref{fig:TRPL_x04}(c,d) in the main text, and Table~\ref{SI:table:parameters_TRPL_OO} and Figures~\ref{SI:fig:MA0.4FA0.6PbI3_PL_vs_power}(c,d)]
are: $\tauRX = 10$ to $50$~ps and $\tauReh = 100$ to $500$~ps, 
see Figure~\ref{SI:fig:MA0.4FA0.6PbI3_PL_vs_power}(e).
The ratio between the exciton and \eh{} populations upon pulse arrival typically varies in the range $I_{\rm X}(0) / I_{\rm eh}(0) = 1$ to $10$. When applying the model to the \MAFAPI{} experimental data, the optical orientation degrees can be anywhere between $0$ and $1$, event though always $\PooX(0) > \Pooeh(0)$.
The \eh{} spin relaxation time $\tauSeh$ is generally between $300$~ps and $1000$~ps, see Figure~\ref{SI:fig:MA0.8FA0.2PbI3_OO_times_vs_power}(a). The exciton spin relaxation time $\tauSX$ rarely can be accurately extracted from the data, but it does not drop significantly below $100$~ps in the regimes studied in this work, see Figure~\ref{SI:fig:MA0.8FA0.2PbI3_OO_times_vs_power}(b).

The model features are benchmarked in Figure~\ref{SI:fig:P_OO_model}. The parameter values chosen in Figure~\ref{SI:fig:P_OO_model} are close to the realistic ones, although we do also use a very short $\tauSX = 20$~ps in Figure~\ref{SI:fig:P_OO_model}(c) to highlight the possibility of non-monotonic optical orientation dynamics, see also Figure~\ref{SI:fig:MA0.8FA0.2PbI3_OO_times_vs_power}(c). Similarly, the exciton to \eh{} intensity ratio $\initPLratio$ never appreciably drops below unity in our experiments [see Figure~\ref{SI:fig:MA0.4FA0.6PbI3_PL_vs_power}(e)], even though this might well be the case at higher excitation density values.

In Figures~\ref{SI:fig:P_OO_model}(a--c), we show how the optical orientation dynamics can become non-monotonic by shortening $\tauSX$ from $500$~ps to $20$~ps, provided the exciton optical orientation becomes smaller than that of the \eh{} pairs while still $I_{\rm X} > I_{\rm eh}$. Generally, this requires  $\tauRX \geq \tauSX$, although the exact condition for the dip to appear depends on the ratio $I_{\rm X}(0) / I_{eh}(0)$ [kept at $10$ in Figures~\ref{SI:fig:P_OO_model}(a--c)]. 
In the experiments reported in this paper, we typically find dynamics cases among those presented in Figures~\ref{SI:fig:P_OO_model}(a,b), i.\,e., $\tauRX \ll \tauSX$, where the dynamics are close to biexponential in shape, though at higher powers we do approach the regime shown in Figure~\ref{SI:fig:P_OO_model}(c), where $\tauSX$ and $\tauRX$ are of the same order of magnitude (see Section~\ref{SI:power_dep}).

In the $\tauRX \ll \tauSX$ regime, the initial decay of $\Poo(t)$ is primarily controlled by the exciton recombination time $\tauRX$, and estimates of $\tauSX$ are difficult. This is illustrated in Figures~\ref{SI:fig:P_OO_model}(d,e), where in both panels we compare the dynamics corresponding to $\tauSX = 100$~ps or $200$~ps, except in Figure~\ref{SI:fig:P_OO_model}(d) where $\tauRX = 50$~ps, while in Figure~\ref{SI:fig:P_OO_model}(e) it is shortened to $\tauRX = 10$~ps.
In Figure~\ref{SI:fig:P_OO_model}(e), the $\Poo(t)$ dynamics resulting from $\tauSX = 100$~ps and $200$~ps can easily be separated due to the significant exciton spin relaxation taking place when $I_{\rm X}(t) \gtrsim I_{\rm eh}(0)$, while in Figure~\ref{SI:fig:P_OO_model}(e), they are virtually identical due to the short $\tauRX$ determining the initial optical orientation dynamics. Here, increasing $\tauSX$ beyond about $100-200$~ps does not lead to noticeable changes in the dynamics due to the exciton recombination dynamics determining the initial decay of $\Poo(t)$. Therefore, in most experiments, one can only expect to extract a lower bound for $\tauSX$.

Finally, in Figure~\ref{SI:fig:P_OO_model}(f), we show the effect of varying the initial ratio of the exciton and \eh{} PL intensity components by two orders of magnitude, from $\initPLratio = 0.1$ to $\initPLratio = 10$.
The shape of the dynamics is governed by the time-dependent intensity ratio $I_{\rm X}(t) / I_{\rm eh}(t)$, which approaches 0 as $t \to \infty$ because $\tauRX \ll \tauReh$.
Therefore, no matter what the initial ratio is, sooner or later the $\Poo(t)$ curve converges to $\Pooeh(t)$, see Eq.~\eqref{SI:eq:P_oo_def}. Starting from a biexponential-like behavior at $\initPLratio = 0.1$ which persists up to $\initPLratio = 1$, we approach the $\initPLratio = 10$ curve which shows a slight bend during the initial decay, similarly to Figure~\ref{SI:fig:P_OO_model}(a).
Because the exciton contribution  continuously stronger for each successive curve, the $\Poo(t)$ signal approaches $\Pooeh(t)$ later and later, despite all the relevant time parameters, such as $\tauRX$ and $\tauSX$, remaining constant.

\newpage

\section{Dependence of recombination and optical orientation dynamics on temperature}
\label{SI:temperature_dep}

\begin{figure*}[hbt]
    \centering
    \includegraphics[]{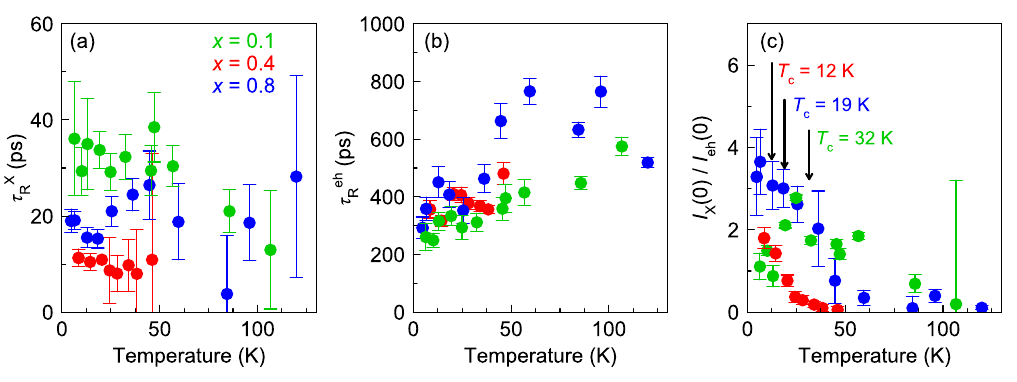}
    \caption{
    (a,b,c) Exciton recombination time, \eh{} recombination time, and initial ratio of exciton and \eh{} PL components plotted as function of temperature. They are measured for three \MAFAPI{} crystals under non-resonant excitation with $E_{\rm exc} = 1.55$~eV, $1.60$~eV, and $1.675$~eV for $x = 0.1$, $0.4$, and $0.8$, respectively, at $P = 0.5$~\WcmSq{}.     The color code for $x$ is identical in these panels.
    In panel (c), the $T_{\rm c}$ values from Figures~\ref{fig:P_OO_temperature}(a--c) are shown by the arrows.
    }
    \label{SI:fig:MAFAPI_PL_vs_T}
\end{figure*}

This section is intended to supplement the temperature-dependent data reported in Section~\ref{sec:temperature} of the main text by providing additional data, notably the temperature dependences of $\tauRX$, $\tauReh$, and $\initPLratio$ in the \MAFAPI{} samples.
They are shown in Figures~\ref{SI:fig:MAFAPI_PL_vs_T}(a,b,c), respectively. As mentioned in the main text, the exciton recombination time $\tauRX$ does not vary appreciably with temperature, remaining approximately between $10$~ps and $40$~ps [see Figure~\ref{SI:fig:MAFAPI_PL_vs_T}(a)] up to $T = 100$~K. 
The carrier recombination time $\tauReh$ tends to become longer with temperature, rising from $200-300$~ps at $T = 5$~K to $500-800$~ps at $T = 100$~K, see Figure~\ref{SI:fig:MAFAPI_PL_vs_T}(b). 
The ratio of exciton and carrier PL components at pulse arrival $\initPLratio$ tends to decrease from $1-5$ at $T = 5$~K to $0$ at higher temperatures [$T \approx 50$~K in \MAFAPIxy{0.4}{0.6} and $T \approx 100$~K for $x=0.1$ and $0.8$, see Figure~\ref{SI:fig:MAFAPI_PL_vs_T}(c)], suggesting that the PL emission in the X spectral range becomes dominated by the recombination of \eh{} pairs at higher temperatures. This type of behavior is generally expected due to thermal ionization of excitons, as the exciton binding energy is about $15$~meV in \MAFAPI{}.
However, it could be that the decrease of the short-lived component in the PL dynamics is governed by the same temperature $T_{\rm c}$ as the decrease of optical orientation degree shown in Figure~\ref{fig:P_OO_temperature}, compare to the $T_{\rm c}$ in Figure~\ref{SI:fig:MAFAPI_PL_vs_T}(c) by the arrows.

\section{Dependence of recombination and optical orientation dynamics on excitation density}
\label{SI:power_dep}

The main results of power-dependent time-resolved optical orientation experiments on the \MAFAPI{} samples are analyzed in Section~\ref{sec:power} of the main text. Here, we present the evolution of the PL and optical orientation dynamics with power in different samples.

\begin{figure*}[t]
    \centering
    \includegraphics[]{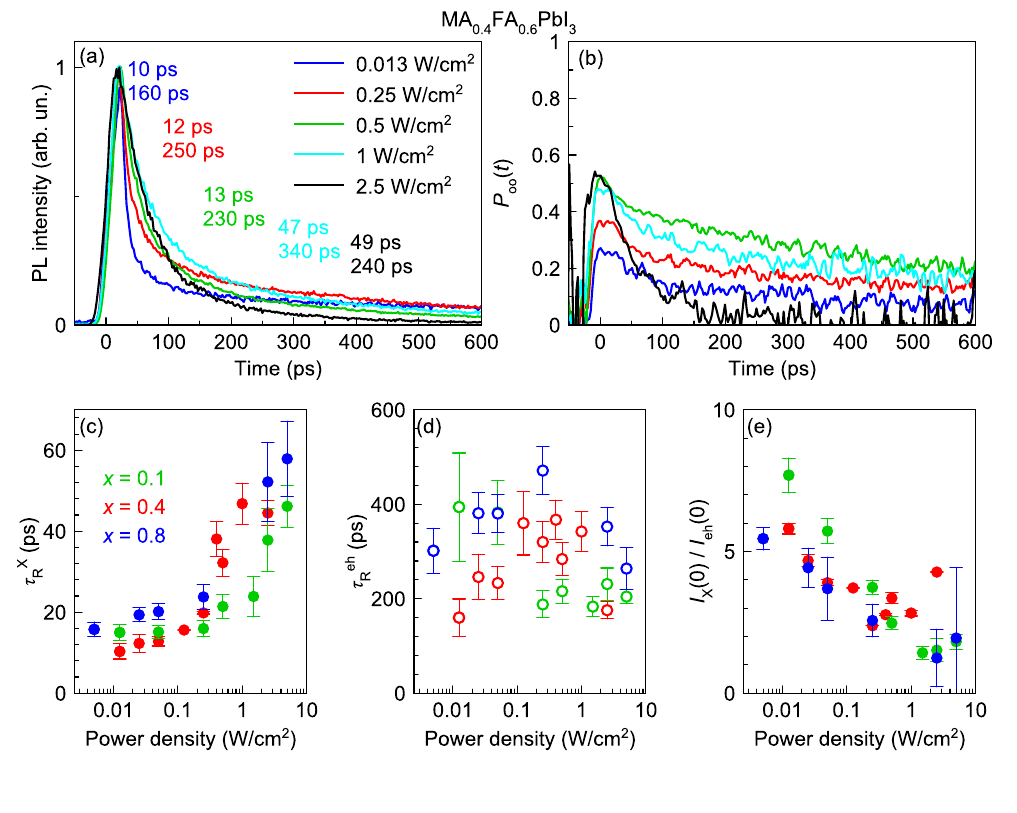}
    \caption{
    Excitation density dependences of the recombination and spin dynamics in \MAFAPI{} crystals measured under non-resonant excitation with $E_{\rm exc} = 1.55$~eV, $1.60$~eV, and $1.675$~eV for $x = 0.1$, $0.4$, and $0.8$, respectively, at $T = 1.6$~K.
    (a) Normalized PL dynamics in \MAFAPIxy{0.4}{0.6} measured at various power densities. 
    (b) Optical orientation dynamics in \MAFAPIxy{0.4}{0.6} measured at various power densities. Colors correspond to the legend in (a).
    (c,d,e) Exciton recombination time, \eh{} recombination time, and initial ratio of exciton and \eh{} PL intensity components as function of excitation density measured on three \MAFAPI{} crystals. The color code for $x$ is the same in these panels.
    }
    \label{SI:fig:MA0.4FA0.6PbI3_PL_vs_power}
\end{figure*}

The excitation density dependence of these dynamics in \MAFAPIxy{0.4}{0.6} are shown in Figures~\ref{SI:fig:MA0.4FA0.6PbI3_PL_vs_power}(a,b). In Figure~\ref{SI:fig:MA0.4FA0.6PbI3_PL_vs_power}(a), the short PL decay time $\tauRX$ becomes longer, while the long decay time $\tauReh$ stays essentially constant with increasing power. In the optical orientation dynamics, the peak value $\Poo(0)$ initially grows with power, then stabilizes. In Figure~\ref{SI:fig:MA0.4FA0.6PbI3_PL_vs_power}(b), the decay of the optical orientation degree initially remains the same, but becomes more pronounced at higher power density values.

The power-dependent behavior of the time-resolved PL exemplified for \MAFAPIxy{0.4}{0.6} in Figures~\ref{SI:fig:MA0.4FA0.6PbI3_PL_vs_power}(a) is common to the other \MAFAPI{} samples, as demonstrated in Figures~\ref{SI:fig:MA0.4FA0.6PbI3_PL_vs_power}(c,d,e). With the power density increasing from $0.01$~\WcmSq{} to $10$~\WcmSq{}, the fast component of the PL decay corresponding to exciton recombination slows from about $10$~ps to about $60$~ps [Figure~\ref{SI:fig:MA0.4FA0.6PbI3_PL_vs_power}(c)]. This is a behavior well-known for semiconductors, when at higher excitation densities the exciton-exciton and the exciton-carrier scattering shorten the exciton coherence time, which results in prolonging the exciton radiative recombination time.
The long decay component corresponding to the recombination of \eh{} pairs remains approximately constant [Figure~\ref{SI:fig:MA0.4FA0.6PbI3_PL_vs_power}(d)], staying in the range of $200-400$~ps. Additionally, the intensity ratio between the exciton and \eh{} recombination components gradually decreases from about $7$ to about $1$, i.\,e., the relative contribution of the carriers to the PL increases with growing power.

\begin{figure*}[t]
    \centering
    \includegraphics[]{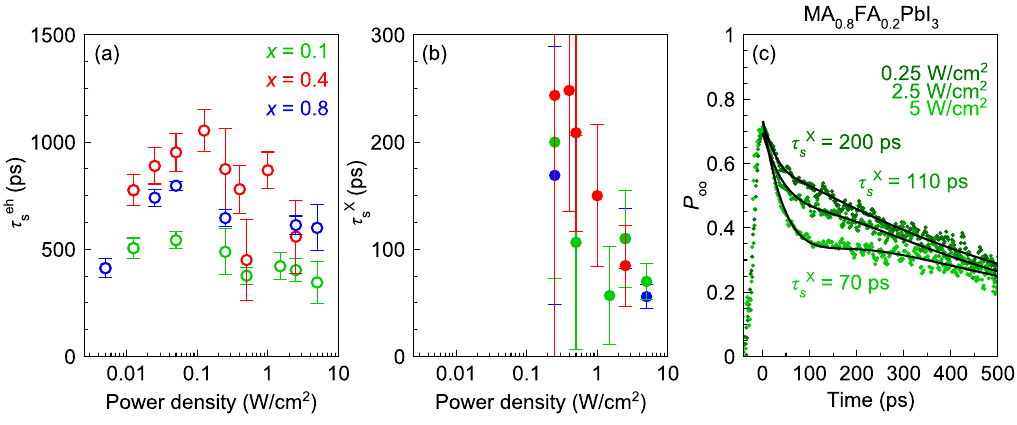}
    \caption{
    (a,b) Carrier and exciton spin relaxation times as function of incident power excitation density, measured under non-resonant excitation with $E_{\rm exc} = 1.55$~eV, $1.60$~eV, and $1.675$~eV for $x = 0.1$, $0.4$, and $0.8$, respectively, at $T = 1.6$~K. $\tauSX$ cannot be determined for $P < 0.25$~\WcmSq{} and is estimated as $\tauSX \gtrsim 300$~ps. 
    (c) Dynamics of the optical orientation degree in a \MAFAPIxy{0.8}{0.2} crystal measured at different excitation densities for non-resonant excitation at $T = 1.6$~K (points). The black lines are fits with Eq.~\eqref{SI:eq:P_oo_def} using the parameters given in Table~\ref{SI:table:parameters_three_powers}.
    }
    \label{SI:fig:MA0.8FA0.2PbI3_OO_times_vs_power}
\end{figure*}

The spin relaxation times extracted by fitting the power-dependent optical orientation dynamics with Eq.~\eqref{SI:eq:P_oo_def} are shown in Figures~\ref{SI:fig:MA0.8FA0.2PbI3_OO_times_vs_power}(a,b). The \eh{} spin relaxation time $\tauSeh$ tends to slightly increase as power rises from $0.01$~\WcmSq{} to $0.1$~\WcmSq{} before slowly dropping as power is increased further. Overall, $\tauSeh$ remains about constant in the regime studied here.
As discussed in Section~\ref{SI:model}, the exciton spin relaxation time $\tauSX$ cannot be extracted when $\tauRX \ll \tauSX$.  However, as $\tauRX$ increases with power, at the higher power densities studied, we can estimate $\tauSX$ with increasing accuracy. These data are presented in Figure~\ref{SI:fig:MA0.8FA0.2PbI3_OO_times_vs_power}(b).
Below $0.25$~\WcmSq{}, $\tauSX$ cannot be determined for any of the samples. Above $0.25$~\WcmSq{}, $\tauSX$ shortens with increasing power, as we approach the regime of non-monotonic optical orientation dynamics described in Section~\ref{SI:model} and shown in Figure~\ref{SI:fig:P_OO_model}(c).
This is illustrated in Figure~\ref{SI:fig:MA0.8FA0.2PbI3_OO_times_vs_power}(c), where the \MAFAPIxy{0.8}{0.2} sample is used as an example for the experimental optical orientation dynamics exhibiting a plateau [intermediate regime between Figures~\ref{SI:fig:P_OO_model}(b,c)].

Overall, for excitation density values between $0.5$~\WcmSq{} and $10$~\WcmSq{}, $\tauSX$ is about $50$ to $150$~ps in all \MAFAPI{} samples.

\begin{table}[hbp]
\renewcommand{\arraystretch}{1.5}
\caption{Parameters used to fit the dynamics of the optical orientation degree measured at various excitation densities on a \MAFAPIxy{0.8}{0.2} crystal at $T=1.6$~K. The fits are shown by the solid lines in Figure~\ref{SI:fig:MA0.8FA0.2PbI3_OO_times_vs_power}(c).}
\begin{ruledtabular}
\begin{tabular}{cccccccc}
$P$ (\WcmSq{}) & \multicolumn{2}{c}{$\Poo(0)$}
                 & \multicolumn{2}{c}{$\tau_{\rm s}$ (ps)}
                 & \multicolumn{2}{c}{$\tau_{\rm R}$ (ps)} 
                 & $\initPLratio$ \\ \hline
     &  X   & $eh$ &  X  & $eh$& X  & $eh$& \\
0.25 & 0.77 & 0.62 & 200 & 650 & 24 & 470 & 2.6 \\
2.5  & 0.77 & 0.60 & 110 & 610 & 52 & 350 & 1.2 \\
5    & 0.89 & 0.55 &  70 & 600 & 58 & 260 & 1.9 \\
\end{tabular}
\end{ruledtabular}
\label{SI:table:parameters_three_powers}
\end{table}



\begin{thebibliography}{33}%

\bibitem{jena2019} A.~K. Jena, A. Kulkarni, T. Miyasaka, 
Halide perovskite photovoltaics: Background, status, and future prospects.
\textit{Chem. Rev.} \textbf{5}, {3036--3103} (2019).

\bibitem{Vinattieri2021_book} \textit{Halide Perovskites for Photonics}, eds.  A. Vinattieri and G. Giorgi,  (AIP Publishing, Melville, New York, \textbf{2021}).

\bibitem{Vardeny2022_book} \textit{Hybrid Organic Inorganic Perovskites: Physical Properties and Applications}, eds. Z. V. Vardeny  and  M. C. Beard, (World Scientific, \textbf{2022}).

\bibitem{protesescu_kovalenko2015nanolett_CsPbX3_bandgap_tuning}
L. Protesescu, S. Yakunin, M. I. Bodnarchuk, F. Krieg, R. Caputo, C. H. Hendon, R. X. Yang, A. Walsh, and M. V. Kovalenko,
\emph{Nanocrystals of cesium lead halide perovskites (CsPbX$_3$, X = Cl, Br, I): Novel optoelectronic materials showing bright emission with wide color gamut},
Nano Lett. \textbf{15}, 3692 (2015). 
\href{https://doi.org/10.1021/nl5048779}{DOI: 10.1021/nl5048779}

\bibitem{wright_herz2016ncomms_FAPbI3Br3_MAPbI3Br3_phonos}
A. D. Wright, C. Verdi, R. L. Milot, G. E. Eperon, M. A. Pérez-Osorio, H. J. Snaith, F. Giustino, M. B. Johnston, and L. M. Herz,
\emph{Electron–phonon coupling in hybrid lead halide perovskites},
Nat. Commun. \textbf{7}, 11755 (2016). \href{https://doi.org/10.1038/ncomms11755}{DOI: 10.1038/ncomms11755}

\bibitem{mohanty_sarma2019acsenlett_FAMAPbI3_phase_diagram}
A. Mohanty, D. Swain, S. Govinda, T. N. G. Row, and D. D. Sarma,
\emph{Phase diagram and dielectric properties of MA$_{1–x}$FA$_{x}$PbI$_3$},
ACS Energy Lett. \textbf{4}, 2045 (2019). \href{https://doi.org/10.1021/acsenergylett.9b01291}{DOI: 10.1021/acsenergylett.9b01291}

\bibitem{lee_park2015advenmat_CsFAPbI3_stable}
J.-W. Lee, D.-H. Kim, H.-S. Kim, S.-W. Seo, S. M. Cho, and N.-G. Park,
\emph{Formamidinium and cesium hybridization for photo- and moisture-stable perovskite solar cell},
Adv. Energy Mater. \textbf{5}, 1501310 (2015). \href{https://doi.org/10.1002/aenm.201501310}{DOI: 10.1002/aenm.201501310}

\bibitem{li_zhu2016chemmat_CsFAPbI3_stable}
Z. Li, M. Yang, J.-S. Park, S.-H. Wei, J. J. Berry, and K. Zhu,
\emph{Stabilizing perovskite structures by tuning tolerance factor: Formation of formamidinium and cesium lead iodide solid-state alloys},
Chem. Mater. \textbf{28}, 284 (2016). \href{https://doi.org/10.1021/acs.chemmater.5b04107}{DOI: 10.1021/acs.chemmater.5b04107}

\bibitem{duan_zhou2018optmat_FAMAPI_PV}
J. Duan, Z. Liu, Y. Zhang, K. Liu, T. He, F. Wang, J. Dai, and P. Zhou,
\emph{Planar perovskite FA$_{x}$MA$_{1-x}$PbI$_3$ solar cell by two-step deposition method in air ambient},
Opt. Mater. \textbf{85}, 55 (2018). \href{https://doi.org/10.1016/j.optmat.2018.07.072}{DOI: 10.1016/j.optmat.2018.07.072}

\bibitem{yang_xiao2019pccp_FAMAPI_PV}
Y. Yang, J. Luo, A. Wei, J. Liu, Y. Zhao, and Z. Xiao,
\emph{Study of perovskite solar cells based on mixed-organic-cation FA$_x$MA$_{1−x}$PbI$_3$ absorption layer},
Phys. Chem. Chem. Phys. \textbf{21}, 11822 (2019). \href{https://doi.org/10.1039/C9CP02003A}{DOI: 10.1039/C9CP02003A}

\bibitem{luo_xiao2020jelmat_FAMAPI_PV}
J. Luo, A. Wei, N. Luo, J. Liu, Y. Zhao, and Z. Xiao,
\emph{Effect of FA$^+$ fraction and dipping time on performance of FA$_x$MA$_{1−x}$PbI$_3$ films and perovskite solar cells},
J. Electron. Mater. \textbf{49}, 7054 (2020). \href{https://doi.org/10.1007/s11664-020-08488-x}{DOI: 10.1007/s11664-020-08488-x}

\bibitem{weber_weller2016jmatchemA_FAMAPbI3_phase_transitions}
O. J. Weber, B. Charles, and M. T. Weller,
\emph{Phase behaviour and composition in the formamidinium–methylammonium hybrid lead iodide perovskite solid solution},
J. Mater. Chem. A \textbf{4}, 15375 (2016). \href{https://doi.org/10.1039/C6TA06607K}{DOI: 10.1039/C6TA06607K}

\bibitem{francisco-lopez_goni2020jphyschemC_FAMAPbI3_phase_diagram}
A. Francisco-López, B. Charles, M. I. Alonso, M. Garriga, M. Campoy-Quiles, M. T. Weller, and A. R. Goñi,
\emph{Phase diagram of methylammonium/formamidinium lead iodide perovskite solid solutions from temperature-dependent photoluminescence and Raman spectroscopies},
J. Phys. Chem. C \textbf{124}, 3448 (2020). \href{https://doi.org/10.1021/acs.jpcc.9b10185}{DOI: 10.1021/acs.jpcc.9b10185}

\bibitem{dyakonov2017spin_ch1}
M. I. Dyakonov, ed.,
\emph{Spin Physics in Semiconductors}, Springer Series in Solid-State Sciences (Springer International, New York, 2017), Chap. 1. \href{https://books.google.fr/books?id=DoZu5QHoHOQC}{Google Books}

\bibitem{opticalOrientation1984}
F. Meier and B. P. Zakharchenya, eds.,
\emph{Optical Orientation} (North-Holland, Amsterdam, 1984).

\bibitem{kopteva_bayer2025PRB_MAPbI3_OO}
N. E. Kopteva, D. R. Yakovlev, E. Yalcin, I. A. Akimov, M. Kotur, B. Turedi, D. N. Dirin, M. V. Kovalenko, and M. Bayer,
\emph{Optical orientation of excitons and charge carriers in methylammonium lead iodide perovskite single crystals in the orthorhombic phase},
Phys. Rev. B \textbf{111}, 195201 (2025). \href{https://doi.org/10.1103/PhysRevB.111.195201}{DOI: 10.1103/PhysRevB.111.195201}

\bibitem{kopteva_bayer2024advSci_FAPbI3_OO}
N. E. Kopteva, D. R. Yakovlev, E. Yalcin, I. A. Akimov, M. O. Nestoklon, M. M. Glazov, M. Kotur, D. Kudlacik, E. A. Zhukov, E. Kirstein, O. Hordiichuk, D. N. Dirin, M. V. Kovalenko, and M. Bayer,
\emph{Highly-polarized emission provided by giant optical orientation of exciton spins in lead halide perovskite crystals},
Adv. Sci. \textbf{11}, 2403691 (2024). \href{https://doi.org/10.1002/advs.202403691}{DOI: 10.1002/advs.202403691}

\bibitem{kudlacik2024optical}
D.~Kudlacik, N.~E.~Kopteva, M.~Kotur, D.~R.~Yakovlev, K.~V.~Kavokin, 
C.~Harkort, M.~Karzel, E.~A.~Zhukov, E.~Evers, V.~V.~Belykh, and M.~Bayer, 
Optical spin orientation of localized electrons and holes interacting with nuclei in a FA$_{0.9}$Cs$_{0.1}$PbI$_{2.8}$Br$_{0.2}$ perovskite crystal,
ACS Photonics \textbf{11}, 2757 (2024).

\bibitem{kotur2025nucleiFAPI}
M. Kotur, P. S. Bazhin, K. V. Kavokin, N. E. Kopteva, D. R. Yakovlev, D. Kudlacik, and M. Bayer, 
Dynamic polarization of nuclear spins by optically-oriented electrons and holes in lead halide perovskite semiconductors,
CondMat ArXive (2025) https://arxiv.org/abs/2509.15530.

\bibitem{dyakonov2017spin}
\textit{Spin Physics in Semiconductors}, edited by M.~I.~Dyakonov (Springer International Publishing AG, 2017).

\bibitem{Grisard2023}
S. Grisard, A. V. Trifonov, I. A. Solovev, D. R. Yakovlev, O. Hordiichuk, M. V. Kovalenko, M. Bayer, and I. A. Akimov, 
Long-lived exciton coherence in mixed-halide perovskite crystals,
Nano Lett. \textbf{23}, 7397 (2023).

\bibitem{galkowski_nicholas2016enenvsci_FAPI_MAPI_Eg_Ex}
K. Galkowski, A. Mitioglu, A. Miyata, P. Plochocka, O. Portugall, G. E. Eperon, J. T.-W. Wang, T. Stergiopoulos, S. D. Stranks, H. J. Snaith, and R. J. Nicholas,
\emph{Determination of the exciton binding energy and effective masses for methylammonium and formamidinium lead tri-halide perovskite semiconductors},
Energy Environ. Sci. \textbf{9}, 962 (2016). \href{https://doi.org/10.1039/C5EE03435C}{DOI: 10.1039/C5EE03435C}

\bibitem{fang_anoniettaLoi2016lightSciAppl_FAPI_film_spectra}
H.-H. Fang, F. Wang, S. Adjokatse, N. Zhao, J. Even, and M. A. Loi,
\emph{Photoexcitation dynamics in solution-processed formamidinium lead iodide perovskite thin films for solar cell applications},
Light: Sci. Appl. \textbf{5}, e16056 (2016). \href{https://doi.org/10.1038/lsa.2016.56}{DOI: 10.1038/lsa.2016.56}

\bibitem{kirstein_bayer2022acsphot_spin_w_nuclei_MAPbI3}
E. Kirstein, D. R. Yakovlev, E. A. Zhukov, J. Höcker, V. Dyakonov, and M. Bayer,
\emph{Spin dynamics of electrons and holes interacting with nuclei in MAPbI$_3$ perovskite single crystals},
ACS Photonics \textbf{9}, 1375 (2022). \href{https://doi.org/10.1021/acsphotonics.2c00096}{DOI: 10.1021/acsphotonics.2c00096}

\bibitem{kirstein_bayer2022natcomms_universal_g_perovskites}
E. Kirstein, D. R. Yakovlev, M. M. Glazov, E. A. Zhukov, D. Kudlacik, I. V. Kalitukha, V. F. Sapega, G. S. Dimitriev, M. A. Semina, M. O. Nestoklon, E. L. Ivchenko, N. E. Kopteva, D. N. Dirin, O. Nazarenko, M. V. Kovalenko, A. Baumann, J. Höcker, V. Dyakonov, and M. Bayer,
\emph{The Landé factors of electrons and holes in lead halide perovskites: universal dependence on the band gap},
Nat. Commun. \textbf{13}, 3062 (2022). \href{https://doi.org/10.1038/s41467-022-30701-0}{DOI: 10.1038/s41467-022-30701-0}

\bibitem{Kopteva_2025OOX} N. E. Kopteva, D. R. Yakovlev, E. Yalcin, I. V. Kalitukha, I. A. Akimov, M. O. Nestoklon, B. Turedi, O. Hordiichuk, D. N. Dirin, M. V. Kovalenko, and M. Bayer,
\emph{Effect of crystal symmetry of lead halide perovskites on the optical orientation of excitons},  
Adv. Sci. \textbf{12}, 2416782 (2025).
 DOI: 10.1002/advs.202416782

 \bibitem{kochereshko_lavallard1998physsolidstate_Pe_Ph_beats}
V. P. Kochereshko, E. L. Ivchenko, D. R. Yakovlev, and F. Lavallard,
\emph{Resonant optical orientation and alignment of excitons in superlattices},
Phys. Solid State \textbf{40}, 2024 (1998). \href{https://doi.org/10.1134/1.1130707}{DOI: 10.1134/1.1130707}

\bibitem{Kirstein_FAPI_2022} E. Kirstein, D. R. Yakovlev, M. M. Glazov, E. Evers, E. A. Zhukov, V. V. Belykh, N. E. Kopteva, D. Kudlacik, O. Nazarenko, D. N. Dirin, M. V. Kovalenko, and M. Bayer,
\emph{Lead-dominated hyperfine interaction impacting the carrier spin dynamics in halide perovskites,}  
Adv. Mater. \textbf{34}, 2105263 (2022).
DOI: 10.1002/adma.202105263

\bibitem{Grisard_2024FAPI} S. Grisard, A. V. Trifonov, T. Hahn, T. Kuhn, O. Hordiichuk, M. V. Kovalenko, D. R. Yakovlev, M. Bayer, and I. A. Akimov,
\emph{Spin-dependent exciton-exciton interactions in a mixed lead halide perovskite crystal},  
ACS Photonics \textbf{11}, 2930-2937 (2024).
https://doi.org/10.1021/acsphotonics.4c00499

\bibitem{chen_bakr2019acsenlett_ITC_MAPbI3_PV}
Z. Chen, B. Turedi, A. Y. Alsalloum, C. Yang, X. Zheng, I. Gereige, A. AlSaggaf, O. F. Mohammed, and O. M. Bakr,
\emph{Single-crystal MAPbI$_3$ perovskite solar cells exceeding 21\% power conversion efficiency},
ACS Energy Lett. \textbf{4}, 1258 (2019). \href{https://doi.org/10.1021/acsenergylett.9b00847}{DOI: 10.1021/acsenergylett.9b00847}

\bibitem{alsalloum_bakr2020acsenlett_ITC_MAPbI3_PV}
A. Y. Alsalloum, B. Turedi, X. Zheng, S. Mitra, A. A. Zhumekenov, K. J. Lee, P. Maity, I. Gereige, A. AlSaggaf, I. S. Roqan, O. F. Mohammed, and O. M. Bakr,
\emph{Low-temperature crystallization enables 21.9\% efficient single-crystal MAPbI$_3$ inverted perovskite solar cells},
ACS Energy Lett. \textbf{5}, 657 (2020). \href{https://doi.org/10.1021/acsenergylett.9b02787}{DOI: 10.1021/acsenergylett.9b02787}

\bibitem{turedi_bakr2022advmat_ITC_PV_long_diff_lengths}
B. Turedi, M. N. Lintangpradipto, O. J. Sandberg, A. Yazmaciyan, G. J. Matt, A. Y. Alsalloum, K. Almasabi, K. Sakhatskyi, S. Yakunin, X. Zheng, R. Naphade, S. Nematulloev, V. Yeddu, D. Baran, A. Armin, M. I. Saidaminov, M. V. Kovalenko, O. F. Mohammed, and O. M. Bakr,
\emph{Single-crystal perovskite solar cells exhibit close to half a millimeter electron-diffusion length},
Adv. Mater. \textbf{34}, 2202390 (2022). \href{https://doi.org/10.1002/adma.202202390}{DOI: 10.1002/adma.202202390}

\bibitem{yang_mohammed2022acsenlett_ITC_xray_hiquality}
C. Yang, J. Yin, H. Li, K. Almasabi, L. Gutiérrez-Arzaluz, I. Gereige, J.-L. Brédas, O. M. Bakr, and O. F. Mohammed,
\emph{Engineering surface orientations for efficient and stable hybrid perovskite single-crystal solar cells},
ACS Energy Lett. \textbf{7}, 1544 (2022). \href{https://doi.org/10.1021/acsenergylett.2c00431}{DOI: 10.1021/acsenergylett.2c00431}


\end{thebibliography}
\end{document}